\documentclass{article}
\usepackage{amssymb}
\usepackage{amsmath}
\usepackage{amsfonts}
\usepackage{hyperref}
\usepackage{graphicx}%
\usepackage{float}
\usepackage{caption}
\begin{document}

\title{Geometric Analogues of Mie\ Scattering}
\author{S. Subedi$^{\infty}$ and T. Curtright$^{\S }$\\Department of Physics, University of Miami, Coral Gables, Florida 33124-8046, USA\\$^{\infty}${\footnotesize sushil.subedi04@gmail.com} $\ \ \ ^{\S }%
${\footnotesize curtright@miami.edu\ }}
\date{}
\maketitle

\begin{abstract}
Cross-sections for particles scattered from selected spatial geometries
exhibit many of the same interesting features as Mie scattering.

\end{abstract}
\tableofcontents
\section{Introduction}

Scattering of light from dielectric and/or conductive spheres (so-called
Lorenz-Mie scattering) is well-known to exhibit fascinating but very
complicated structure \cite{Hulst}. \ There are numerous poles in the various
scattering amplitudes, resulting in very narrow as well as rather broad
resonances, and for resistive spheres there can be substantial power loss,
i.e. inelastic scattering. \ 

Similar effects are easily produced by spherical scattering centers consisting
of nontrivial spatial geometries and nothing else \cite{CS,S,CSunpub}. \ A
detailed discussion of selected examples involving such alternative scattering
mechanisms is the subject of this paper. \ The analysis is carried out in the
context of the three-dimensional Helmholtz equation wherein only spatial
geometry is modified from Euclidean space. \ Time is taken to be universal to
all frames. \ In this sense the study is non-relativistic.

There are known equivalence relations for scattering by potentials, by
dielectric materials, and by various geometric features \cite{Equiv,Dialog}.
\ Such equivalences are not invoked here. \ Rather, the emphasis here is on
the geometry and the resulting cross-sections as computed directly from
partial wave expansions. \ Nevertheless, as a practical matter, the analysis
to follow may provide insight into the behavior of mathematically equivalent
meta-materials. \ Alternatively, the following may provide useful intuition
about scattering phenomena in general relativity, at least in the
non-relativistic limit. \ There are some previous investigations along these
same lines \cite{M1,M2}, including studies of the corresponding Born series,
which place emphasis on other aspects of the same subject.

Relevant formulae are given in the body of the paper along with selected
numerical graphs to illustrate various important physical effects.
\ Additional plots are collected in an Appendix.

\newpage

\section{Scattering by spheres}

A brief review of scattering by a homogeneous sphere of radius $R$ embedded in
three-dimensional Euclidean space is warranted for purposes of comparison.
\ Scattering of plane waves by impenetrable \textquotedblleft
hard\textquotedblright\ spheres in non-relativistic quantum mechanics and\ by
perfectly conducting spheres in classical electromagnetic theory are described
by the following integrated cross-sections \cite{Newton,Jackson,Zangwill}.%
\begin{equation}
\sigma_{\text{QM hard sphere}}=\sum_{l=0}^{\infty}\sigma_{l\text{ QM hard
sphere}}\ ,\ \ \ \sigma_{l\text{ QM hard sphere}}=\frac{4\pi}{k^{2}}\left(
2l+1\right)  \sin^{2}\left(  \delta_{l}\right)
\end{equation}%
\begin{equation}
\sigma_{\text{EM perf cond sphere}}=\sigma_{\text{TE perf cond sphere}}%
+\sigma_{\text{TM perf cond sphere}}%
\end{equation}%
\begin{align}
\sigma_{\text{TE perf cond sphere}}  &  =\sum_{l=1}^{\infty}\sigma_{l\text{ TE
perf}}\ ,\ \ \ \sigma_{l\text{ TE perf}}=\frac{2\pi}{k^{2}}\left(
2l+1\right)  \sin^{2}\left(  \delta_{l}\right) \\
\sigma_{\text{TM perf cond sphere}}  &  =\sum_{l=1}^{\infty}\sigma_{l\text{ TM
perf}}\ ,\ \ \ \sigma_{l\text{ TM perf}}=\frac{2\pi}{k^{2}}\left(
2l+1\right)  \sin^{2}\left(  \Delta_{l}\right)
\end{align}%
\begin{equation}
e^{2i\delta_{l}}=-h_{l}^{\left(  2\right)  }\left(  kR\right)  /h_{l}^{\left(
1\right)  }\left(  kR\right)  \ ,\ \ \ e^{2i\Delta_{l}}=-\frac{d}{dR}\left(
Rh_{l}^{\left(  2\right)  }\left(  kR\right)  \right)  /\frac{d}{dR}\left(
Rh_{l}^{\left(  1\right)  }\left(  kR\right)  \right)
\end{equation}
The scattering is purely elastic with real phase shifts $\delta_{l}$ and
$\Delta_{l}$. \ With the use of the Bessel function relations%
\begin{align}
h_{l}^{\left(  1,2\right)  }\left(  x\right)   &  =\sqrt{\frac{\pi}{2x}%
}\left(  J_{l+\frac{1}{2}}\left(  x\right)  \pm iY_{l+\frac{1}{2}}\left(
x\right)  \right) \\
\frac{d}{dx}\left(  \sqrt{x}J_{l+1/2}\left(  x\right)  \right)   &  =\frac
{1}{\sqrt{x}}\left(  \left(  l+1\right)  J_{l+\frac{1}{2}}\left(  x\right)
-xJ_{l+\frac{3}{2}}\left(  x\right)  \right) \\
\frac{d}{dx}\left(  \sqrt{x}Y_{l+1/2}\left(  x\right)  \right)   &  =\frac
{1}{\sqrt{x}}\left(  \left(  l+1\right)  Y_{l+\frac{1}{2}}\left(  x\right)
-xY_{l+\frac{3}{2}}\left(  x\right)  \right)
\end{align}
it follows that%
\begin{equation}
\frac{1}{\pi R^{2}}~\sigma_{\text{QM hard sphere}}=\frac{4}{x^{2}}\sum
_{l=0}^{\infty}\left(  2l+1\right)  \frac{\left(  J_{l+1/2}\left(  x\right)
\right)  ^{2}}{\left(  J_{l+1/2}\left(  x\right)  \right)  ^{2}+\left(
Y_{l+1/2}\left(  x\right)  \right)  ^{2}}%
\end{equation}%
\begin{equation}
\frac{1}{\pi R^{2}}~\sigma_{\text{TE perf cond sphere}}=\frac{2}{x^{2}}%
\sum_{l=1}^{\infty}\left(  2l+1\right)  \frac{\left(  J_{l+1/2}\left(
x\right)  \right)  ^{2}}{\left(  J_{l+1/2}\left(  x\right)  \right)
^{2}+\left(  Y_{l+1/2}\left(  x\right)  \right)  ^{2}}=\frac{1}{2}%
~\frac{\sigma_{\text{QM hard sphere}}}{\pi R^{2}}%
\end{equation}%
\begin{equation}
\frac{1}{\pi R^{2}}~\sigma_{\text{TM perf cond sphere}}=\frac{2}{x^{2}}%
\sum_{l=1}^{\infty}\left(  2l+1\right)  \frac{\left(  \left(  l+1\right)
J_{l+\frac{1}{2}}\left(  x\right)  -xJ_{l+\frac{3}{2}}\left(  x\right)
\right)  ^{2}}{\left(  \left(  l+1\right)  J_{l+\frac{1}{2}}\left(  x\right)
-xJ_{l+\frac{3}{2}}\left(  x\right)  \right)  ^{2}+\left(  \left(  l+1\right)
Y_{l+\frac{1}{2}}\left(  x\right)  -xY_{l+\frac{3}{2}}\left(  x\right)
\right)  ^{2}}%
\end{equation}
where $x\equiv kR$. \ For these examples, there are no fields inside the
sphere. \ 

For dielectric spheres, on the other hand, the electric and magnetic fields
are non-zero within the sphere. \ With\ $\mu=\mu_{0}$ and real, constant index
of refraction inside the sphere, $n^{2}=\varepsilon/\varepsilon_{0}$, the
standard electromagnetic boundary conditions yield cross-sections that are
given by the following partial wave expansions .%
\begin{equation}
\frac{1}{\pi R^{2}}~\sigma_{\text{dielectric sphere TE + TM}}=\frac{2}{x^{2}%
}\sum_{l=1}^{\infty}\left(  2l+1\right)  \left(  \left\vert \frac{S_{l}%
^{TE}\left(  x\right)  -1}{2i}\right\vert ^{2}+\left\vert \frac{S_{l}%
^{TM}\left(  x\right)  -1}{2i}\right\vert ^{2}\right)  \text{ , \ \ }x\equiv
kR
\end{equation}%
\begin{equation}
S_{l}^{TE}\left(  kR\right)  =-\frac{j_{l}\left(  nkR\right)  \frac{d}%
{dR}\left(  R~h_{l}^{\left(  2\right)  }\left(  kR\right)  \right)
-h_{l}^{\left(  2\right)  }\left(  kR\right)  \frac{d}{dR}\left(
R~j_{l}\left(  nkR\right)  \right)  }{j_{l}\left(  nkR\right)  \frac{d}%
{dR}\left(  R~h_{l}^{\left(  1\right)  }\left(  kR\right)  \right)
-h_{l}^{\left(  1\right)  }\left(  kR\right)  \frac{d}{dR}\left(
R~j_{l}\left(  nkR\right)  \right)  }%
\end{equation}%
\begin{equation}
S_{l}^{TM}\left(  kR\right)  =-\frac{h_{l}^{\left(  2\right)  }\left(
kR\right)  \frac{d}{dR}\left(  R~j_{l}\left(  nkR\right)  \right)  -n^{2}%
j_{l}\left(  nkR\right)  \frac{d}{dR}\left(  R~h_{l}^{\left(  2\right)
}\left(  kR\right)  \right)  }{h_{l}^{\left(  1\right)  }\left(  kR\right)
\frac{d}{dR}\left(  R~j_{l}\left(  nkR\right)  \right)  -n^{2}j_{l}\left(
nkR\right)  \frac{d}{dR}\left(  R~h_{l}^{\left(  1\right)  }\left(  kR\right)
\right)  }%
\end{equation}
For real $n$ the scattering is once again purely elastic with $\left\vert
S_{l}^{TE}\right\vert =1=\left\vert S_{l}^{TM}\right\vert $ for all $k$, and
hence $S_{l}^{TE}=\exp\left(  2i\delta_{l}\right)  $ and $S_{l}^{TM}%
=\exp\left(  2i\Delta_{l}\right)  $ with real phase shifts $\delta_{l}$ and
$\Delta_{l}$, albeit now given by expressions more complicated than the
perfectly conducting case. \ For non-resistive material with real $n$, there
is no electromagnetic power loss, i.e. no energy absorption within the sphere.
\ However, energy absorption does take place if the material within the sphere
is conductive with resistivity $\rho$.

The simplest model for resistive material is obtained by invoking a linear
Ohm's law relation within the sphere, $\overrightarrow{J}=\overrightarrow{E}%
/\rho$. \ If in addition $\varepsilon\neq\varepsilon_{0}$ but $\mu=\mu_{0}$,
then for monochromatic waves \
\begin{equation}
n^{2}=\frac{1}{\varepsilon_{0}}\left(  \varepsilon+\frac{i}{\omega\rho
}\right)  \ ,\ \ \ \omega=kc\
\end{equation}
The previous results then hold with the substitution $\varepsilon
\rightarrow\varepsilon+i/\left(  \omega\rho\right)  $. \ Note the SI units
$\left[  \rho\right]  =[Ohm~Meter]$. \ So $\rho/\left(  Z_{0}R\right)  $ and
$\varepsilon_{0}\omega\rho=x\rho/\left(  Z_{0}R\right)  $ are dimensionless
numbers, with $Z_{0}=\sqrt{\mu_{0}/\varepsilon_{0}}\approx377~Ohms$ and
dimensionless $x=kR$. \ For this simple model, an obvious method to obtain
numerical results approaching those for a perfectly conducting sphere is to
take decreasing values for the resistivity $\rho$. \ 

For non-relativistic QM scattering, effects similar to those for non-resistive
dielectric spheres can be obtained by considering finite, real, constant
potentials within the sphere \cite{Newton}. \ By making the potential complex
valued, it is also possible to exhibit QM scattering effects similar to those
produced by a conductive dielectric sphere with non-zero resistivity. \ 

A more detailed discussion of this QM model is helpful for the remainder of
the paper. \ Outside the sphere, $V=0$, and the stationary state wave function
with energy $\hbar^{2}k^{2}/\left(  2m\right)  $ is given by the partial wave
expansion,
\begin{equation}
\psi\left(  r,\theta\right)  =\sum_{l=0}^{\infty}i^{l}~\frac{1}{2}\left(
S_{l}~h_{l}^{\left(  1\right)  }\left(  kr\right)  +h_{l}^{\left(  2\right)
}\left(  kr\right)  \right)  \left(  2l+1\right)  P_{l}\left(  \cos
\theta\right)  \text{ \ \ for \ \ }R\leq r\leq\infty
\end{equation}
The scattering amplitude, the differential and integrated \emph{elastic}
cross-sections, and the \emph{total} cross-section are then given by the
general relations,%
\begin{align}
f\left(  \theta\right)   &  =\frac{1}{k}\sum_{l=0}^{\infty}\left(
2l+1\right)  \left(  \frac{S_{l}-1}{2i}\right)  P_{l}\left(  \cos
\theta\right)  \ ,\ \ \ \frac{d\sigma_{el}}{d\Omega}=\left\vert f\left(
\theta\right)  \right\vert ^{2}\ ,\ \ \ \sigma_{tot}=\sigma_{el}+\sigma
_{inel}\\
\sigma_{el}  &  =\int\frac{d\sigma_{el}}{d\Omega}~d\Omega=\frac{4\pi}{k^{2}%
}\sum_{l=0}^{\infty}\left(  2l+1\right)  \left\vert \frac{S_{l}-1}%
{2i}\right\vert ^{2}\ ,\ \ \ \sigma_{tot}=\frac{2\pi}{k^{2}}\sum_{l=0}%
^{\infty}\left(  2l+1\right)  ~\left(  1-\operatorname{Re}S_{l}\right)
\end{align}
\ where the \emph{inelastic} cross-section is defined as $\sigma_{inel}%
=\sigma_{tot}-\sigma_{el}=\frac{\pi}{k^{2}}\sum_{l=0}^{\infty}\left(
2l+1\right)  \left(  1-\left\vert S_{l}\right\vert ^{2}\right)  $.

Inside the sphere, let%
\begin{equation}
V\left(  r\right)  =\frac{\hbar^{2}\eta}{2m}\text{ \ \ for \ \ }0\leq r\leq R
\end{equation}
Define $x\equiv kR$, $u\equiv\eta R^{2}$, and $\kappa^{2}\equiv k^{2}-\eta$,
so $\kappa R=\sqrt{x^{2}-u}$. \ Continuity of $\psi$ and it's first
derivatives at $r=R$ leads to%
\begin{equation}
S_{l}=-\left(  \frac{x~J_{l+1/2}\left(  \sqrt{x^{2}-u}\right)  H_{l+3/2}%
^{\left(  2\right)  }\left(  x\right)  -\sqrt{x^{2}-u}~J_{l+3/2}\left(
\sqrt{x^{2}-u}\right)  H_{l+1/2}^{\left(  2\right)  }\left(  x\right)
}{x~J_{l+1/2}\left(  \sqrt{x^{2}-u}\right)  H_{l+3/2}^{\left(  1\right)
}\left(  x\right)  -\sqrt{x^{2}-u}~J_{l+3/2}\left(  \sqrt{x^{2}-u}\right)
H_{l+1/2}^{\left(  1\right)  }\left(  x\right)  }\right)
\end{equation}
The coefficients appearing in $f\left(  \theta\right)  $ and $\sigma_{el}$ are
then%
\begin{equation}
A_{l}=\frac{i}{2}\left(  1-S_{l}\right)  =i\left(  \frac{x~J_{l+1/2}\left(
\sqrt{x^{2}-u}\right)  J_{l+3/2}\left(  x\right)  -\sqrt{x^{2}-u}%
~J_{l+3/2}\left(  \sqrt{x^{2}-u}\right)  J_{l+1/2}\left(  x\right)
}{x~J_{l+1/2}\left(  \sqrt{x^{2}-u}\right)  H_{l+3/2}^{\left(  1\right)
}\left(  x\right)  -\sqrt{x^{2}-u}~J_{l+3/2}\left(  \sqrt{x^{2}-u}\right)
H_{l+1/2}^{\left(  1\right)  }\left(  x\right)  }\right)
\end{equation}
For real $\eta$ (i.e. real $u=\eta R^{2}$) it follows that $\left\vert
S_{l}\right\vert =1$ for all real $k$ (i.e. real $x=kR$). \ In this case
$\sigma_{inel}=0$. \ But in general, for complex $\eta$ (i.e. complex $u$),
$\left\vert S_{l}\right\vert \neq1$ and $\sigma_{inel}\neq0$.

\section{Selected spatial geometries}

A simple \textquotedblleft foxhole\textquotedblright\ geometric model that is
able to mimic features of Mie scattering for real index of refraction is given
by endowing a three dimensional manifold with a non-trivial metric as obtained
by the following $\mathbb{M}_{3}\subset\mathbb{E}_{4}$ embedding.
\begin{gather}
\left(  ds\right)  ^{2}=\left(  dh\right)  ^{2}+\left(  dr\right)  ^{2}%
+r^{2}\left(  d\theta\right)  ^{2}+\left(  r^{2}\sin^{2}\theta\right)  \left(
d\phi\right)  ^{2}\\
h\left(  r,n\right)  =\frac{-H}{\left(  1+\left(  r/R\right)  ^{2n}\right)
^{p}}\ ,\ \ \ 0\leq r\leq\infty
\end{gather}
where $\theta$ and $\phi$ are the usual spherical polar angles in
$\mathbb{E}_{3}$. \ For example, when $p=1/2$, equatorial slices of the
manifold for any fixed $\phi$, for various $n$, are pictured here:%
\begin{figure}[htb]
	\centering
	\includegraphics[width=5in]{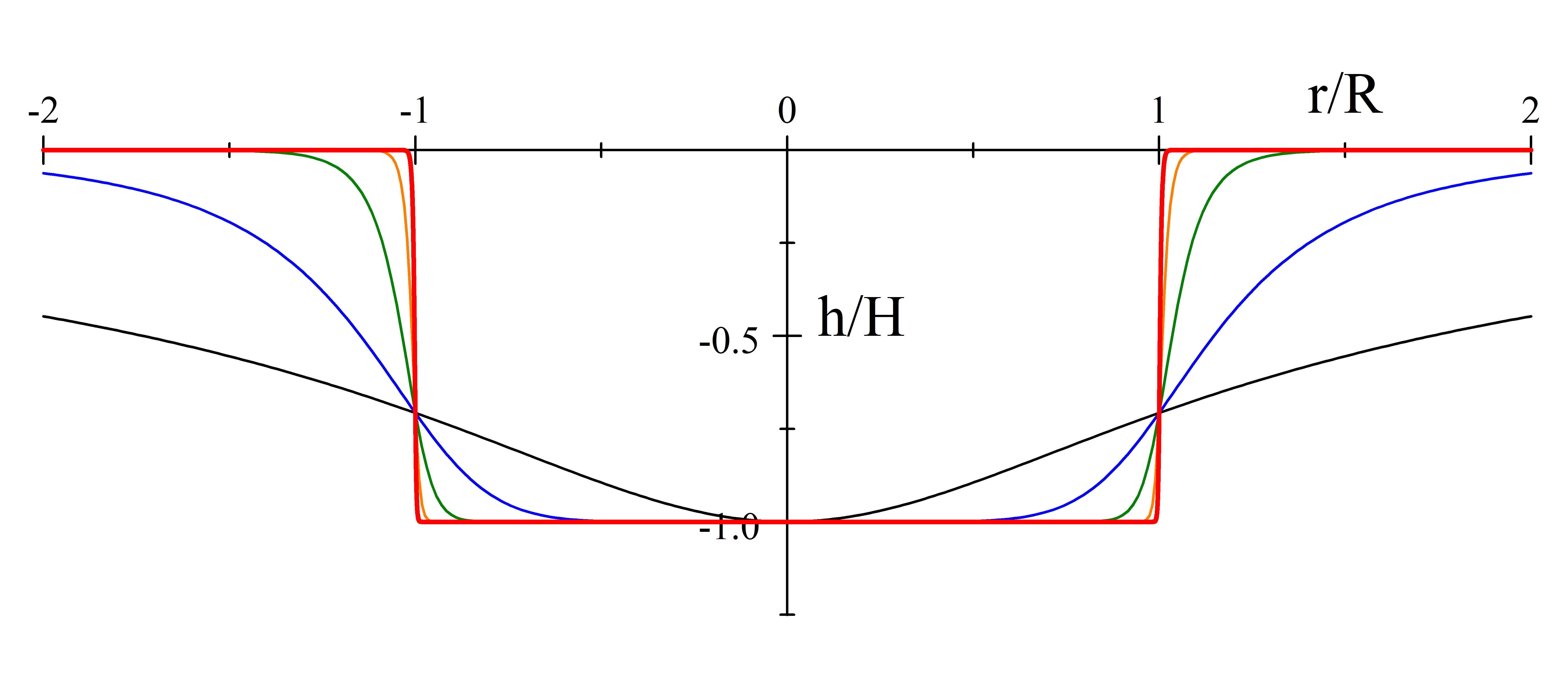} 
	\caption{ $\ \frac{h}{H}=\frac{-1}{\sqrt{1+\left(  r/R\right)  ^{2n}}}$ for
		$n=1,\ 4,\ 16,\ 64$ \& $256$.}
\end{figure}

Analytically, the limit $n\rightarrow\infty$ is somewhat obscure, but from the
previous graph, that limit is clear. \ As $n\rightarrow\infty$ the geometry is
a cylinder of height $H$ composed of 2-spheres each of radius $R$ (i.e.
$\left[  -H,0\right]  \otimes S_{2}\left(  R\right)  $) terminated at the
\textquotedblleft lower\textquotedblright\ end by a ball of radius $R$ (i.e.
$\mathbb{B}_{3}\left(  R\right)  $) and attached at the \textquotedblleft
upper\textquotedblright\ end to a punctured $\mathbb{E}_{3}$ (i.e.
$\mathbb{E}_{3}\smallsetminus\mathbb{B}_{3}\left(  R\right)  $). \ It will
turn out that scattering from this foxhole\ geometry is elastic for real $H$,
but if $H$ is taken to be complex, the model can produce inelastic scattering.

Alternatively, a geometric model obtained from a wormhole metric also exhibits
features of Mie scattering with absorption. \ Consider a smooth, spatial
\textquotedblleft bridge\textquotedblright\ manifold \cite{EandR} (i.e. a
static \textquotedblleft wormhole\textquotedblright\ \cite{Thorne}) defined by%
\begin{align}
\left(  ds\right)  ^{2}  &  =R^{2}\left(  dw\right)  ^{2}+r^{2}\left(
w\right)  \left(  d\theta\right)  ^{2}+\left(  r^{2}\left(  w\right)  \sin
^{2}\theta\right)  \left(  d\phi\right)  ^{2}\\
r\left(  w\right)   &  =R\left(  1+\left(  w^{2}\right)  ^{n/2}\right)
^{1/n}\ ,\ \ \ -\infty\leq w\leq+\infty
\end{align}
For example, the static version of the Ellis metric \cite{Ellis} is given by
$n=2$, namely,
\[
r\left(  w\right)  =R\sqrt{1+w^{2}}\ ,\ \ \ -\infty\leq w\leq+\infty
\]%
\begin{equation}
x\left(  w,\theta\right)  =r\left(  w\right)  \cos\theta\ ,\ \ \ y\left(
w,\theta\right)  =r\left(  w\right)  \sin\theta
\end{equation}
As before, a 2D equatorial slice of the manifold (i.e. $\theta=\pi/2$, hence
$z=0$) can be embedded in 3D, only now with the help of an \textquotedblleft
extra-physical\textquotedblright\ dimension, $h$. \ On that slice \
\begin{equation}
\left(  ds\right)  ^{2}=R^{2}\left(  dw\right)  ^{2}+r^{2}\left(  w\right)
\left(  d\phi\right)  ^{2}=\left(  dx\right)  ^{2}+\left(  dy\right)
^{2}+\frac{1}{\alpha^{2}}\left(  dh\right)  ^{2}%
\end{equation}%
\begin{equation}
x\left(  w,\phi\right)  =r\left(  w\right)  \cos\phi\ ,\ \ \ y\left(
w,\phi\right)  =r\left(  w\right)  \sin\phi
\end{equation}%
\begin{equation}
h\left(  w\right)  =\alpha\int_{0}^{w}\sqrt{R^{2}-\left(  dr\left(
\varpi\right)  /d\varpi\right)  ^{2}}d\varpi=\alpha R\int_{0}^{w}%
\sqrt{1-\left(  \varpi^{2}\right)  ^{n-1}\left(  1+\left(  \varpi^{2}\right)
^{n/2}\right)  ^{\frac{2}{n}-2}}\,d\varpi
\end{equation}
where $\alpha\equiv H/R$ is the \textquotedblleft aspect
ratio\textquotedblright\ of the wormhole. \ For example, for the Ellis case
with $\alpha=1$ and $n=2$,%
\begin{equation}
h\left(  w\right)  =R\ln\left(  w+\sqrt{1+w^{2}}\right)
=R\operatorname{arcsinh}\left(  w\right)
\end{equation}
For generic $n$, it is easiest to obtain $h\left(  w\right)  $ by numerical
solution of
\begin{equation}
\frac{1}{R}\frac{dh\left(  w\right)  }{dw}=\alpha\sqrt{1-\left(  w^{2}\right)
^{n-1}\left(  1+\left(  w^{2}\right)  ^{n/2}\right)  ^{\frac{2}{n}-2}}%
\end{equation}
with initial condition $h\left(  0\right)  =0$. \ 

Here are equatorial slice profiles for the $\left(  x,h\right)  $ plane of the
embedded manifolds, for five different values of $n$. \ The same profile is
obtained for the $\left(  y,h\right)  $ plane, or for any other plane oriented
at a fixed azimuthal angle. \ That is to say, the embedding is rotationally
invariant about the $h$ axis.%
\begin{figure}[htb]
	\centering
	\includegraphics[width=5in]{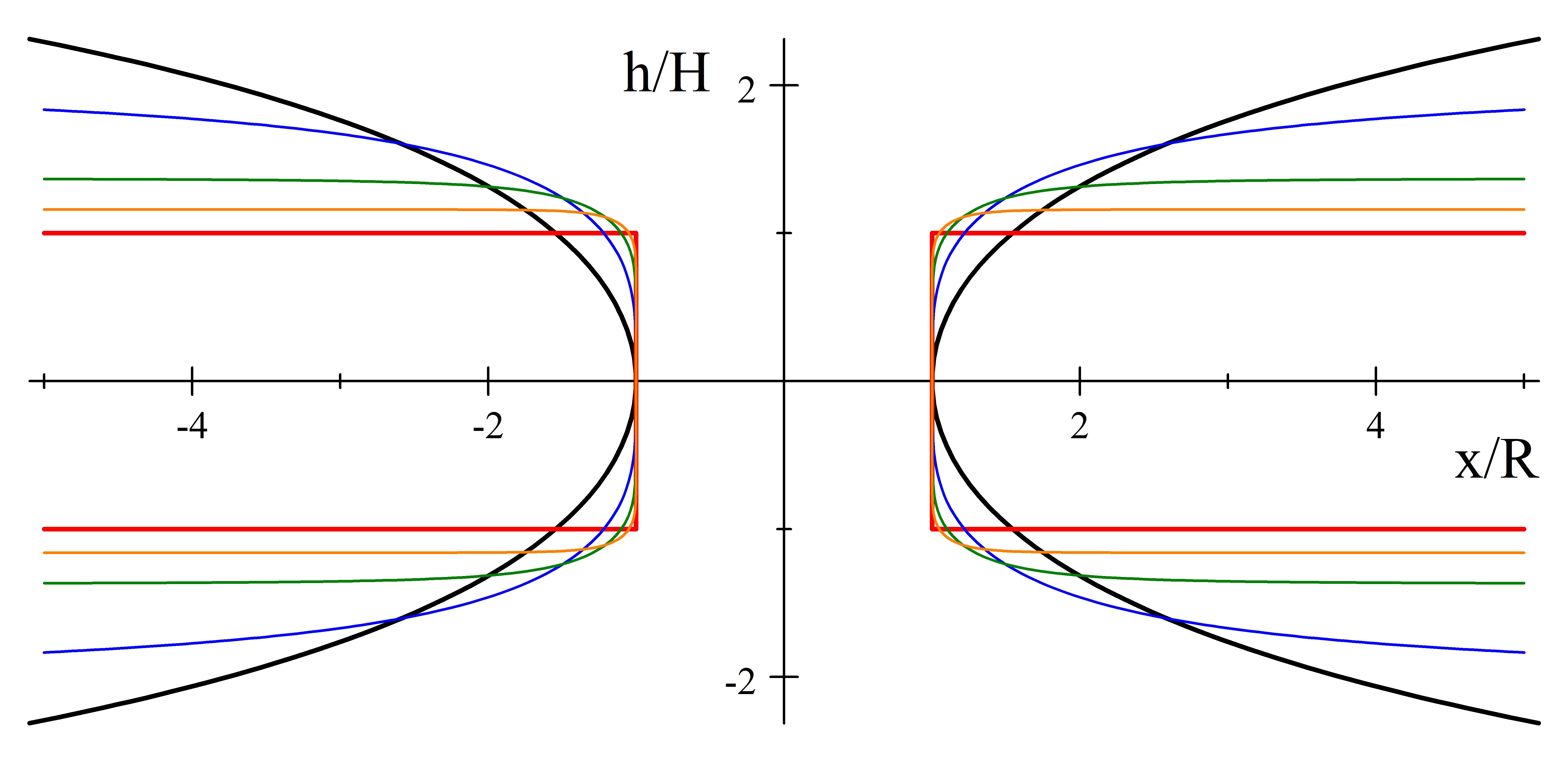} 
	\caption{Equatorial slice profiles for $n=2,\ 4,\ 8,\ 16,$ \& $\infty$    in black, blue, green, orange, and red, respectively.}
\end{figure}

Once again, the limit $n\rightarrow\infty$ is somewhat difficult to process
analytically, but from the previous graph, that limit is again clear. \ As
$n\rightarrow\infty$ the geometry is a cylinder of height $2H$ composed of
2-spheres each of radius $R$ (i.e. $\left[  -H,H\right]  \otimes S_{2}\left(
R\right)  $) attached at both ends to a disjoint pair of punctured
$\mathbb{E}_{3}$s (i.e. $\mathbb{E}_{3}\smallsetminus\mathbb{B}_{3}\left(
R\right)  $ at $h=\pm H=\pm\alpha R$).

It so happens that scattering from this wormhole geometry is inelastic for any
$n\geq2$, in the sense that the wormhole absorbs particle flux incident upon
it from infinite distance, say, on the upper branch of the manifold. \ In the
rest of this paper we will consider mostly the $n=\infty$ foxhole and wormhole
geometries, with just a few related remarks for the $n=2$ Ellis wormhole.

Variants of these foxhole and wormhole geometries are easily imagined. \ One
such variant would be to join the finite $n$ geometries smoothly with flat
space at a fixed radius $R_{1}>R$. \ For example, with Heaviside step function
$\Theta$, a modified Ellis wormhole is given by%
\begin{equation}
r\left(  w\right)  =R\left(  1+\left(  w^{2}\right)  ^{n/2}\right)
^{1/n}\Theta\left(  w_{1}^{2}-w^{2}\right)  +R_{1}\sqrt{w^{2}/w_{1}^{2}%
}~\Theta\left(  w^{2}-w_{1}^{2}\right)  ,\ \ \ -\infty\leq w\leq+\infty
\end{equation}
where $R_{1}=R\left(  1+\left(  w_{1}^{2}\right)  ^{n/2}\right)  ^{1/n}$ for
some chosen $w_{1}>0$. \ This modification would give a continuous metric
whose first derivatives are discontinuous at $R_{1}$, and whose second
derivatives are infinite (i.e. Dirac delta terms in the curvature) at $R_{1}$.
\ (This is also true for either of the $n\rightarrow\infty$ manifolds at
$r=R$.) \ However, for the models to be considered the wave function $\Psi$ is
taken to be a solution to the covariant Helmholtz equation,
\begin{equation}
\left(  \nabla^{2}+k^{2}\right)  \Psi=0
\end{equation}
where $\nabla^{2}$ is the invariant Laplacian on the manifold, $\nabla^{2}%
\Psi=\frac{1}{\sqrt{g}}\partial_{\mu}\left(  \sqrt{g}g^{\mu\nu}\partial_{\nu
}\Psi\right)  $. \ Therefore, without coupling $\Psi$ directly to the
curvature, $\Psi$ and its first derivatives are allowed to be continuous at
$R=R_{1}$, so there will be no Dirac deltas in this covariant Helmholtz
equation. \ (This is also true for either of the $n\rightarrow\infty$
manifolds at $r=R$.)

\section{Amplitudes}

The $n\rightarrow\infty$ foxhole geometry consists of three flat space parts:
\ The punctured space $\mathbb{E}_{3}\smallsetminus\mathbb{B}_{3}\left(
R\right)  $, the cylinder composed of $S_{2}\left(  R\right)  $s, and the ball
$\mathbb{B}_{3}\left(  R\right)  $ at $h=-H$. \ On $\mathbb{E}_{3}%
\smallsetminus\mathbb{B}_{3}\left(  R\right)  $ and on $\mathbb{B}_{3}\left(
R\right)  $,
\begin{equation}
\nabla^{2}\Psi=\frac{1}{r}\frac{\partial^{2}}{\partial r^{2}}\left(
r\Psi\right)  -\frac{1}{r^{2}}~L^{2}\Psi
\end{equation}
where $\overrightarrow{L}=-i\overrightarrow{r}\times\overrightarrow{\nabla} $.
\ Upon choosing angular momentum eigenstates, and factoring $\Psi=\psi\left(
r\right)  Y_{lm}\left(  \theta,\phi\right)  $, the radial function must
satisfy%
\begin{equation}
\frac{1}{r}\frac{\partial^{2}}{\partial r^{2}}\left(  r\psi\right)  +\left(
k^{2}-\frac{l\left(  l+1\right)  }{r^{2}}\right)  \psi=0
\end{equation}
with spherical Bessel function solutions $\psi=j_{l}\left(  kr\right)
=\sqrt{\frac{\pi}{2kr}}J_{l+1/2}\left(  kr\right)  $ or $\psi=y_{l}\left(
kr\right)  =\sqrt{\frac{\pi}{2kr}}Y_{l+1/2}\left(  kr\right)  $. \ On the
cylinder%
\begin{equation}
\nabla^{2}\Psi=\frac{\partial^{2}}{\partial h^{2}}\Psi-\frac{1}{R^{2}}%
~L^{2}\Psi
\end{equation}
with definite angular momentum states given by $\Psi=\psi\left(  h\right)
Y_{lm}\left(  \theta,\phi\right)  $. Helmholtz's equation is now satisfied
provided
\begin{equation}
\frac{\partial^{2}}{\partial h^{2}}\psi\left(  h\right)  +\left(  k^{2}%
-\frac{l\left(  l+1\right)  }{R^{2}}\right)  \psi\left(  h\right)
\end{equation}
with solutions $\psi=\sin\left(  \kappa h\right)  $ and $\psi=\cos\left(
\kappa h\right)  $ where $\kappa^{2}=k^{2}-l\left(  l+1\right)  /R^{2}$. \ 

For an incident plane wave $\exp\left(  ikr\cos\theta\right)  $ only the
$\phi$-independent $P_{l}\left(  \cos\theta\right)  =\sqrt{\frac{4\pi}{2l+1}%
}Y_{l0}\left(  \theta,\phi\right)  $ are involved in the scattering solutions.
\ In this case the solutions on $\mathbb{E}_{3}\smallsetminus\mathbb{B}%
_{3}\left(  R\right)  $ take the familiar form \
\begin{equation}
\Psi\left(  r,\theta\right)  =\sum_{l=0}^{\infty}i^{l}~\frac{1}{2}\left(
S_{l}~h_{l}^{\left(  1\right)  }\left(  kr\right)  +h_{l}^{\left(  2\right)
}\left(  kr\right)  \right)  \left(  2l+1\right)  P_{l}\left(  \cos
\theta\right)  \text{ \ \ for \ \ }R<r\leq\infty
\end{equation}
with coefficients $S_{l}$ to be determined, while on the ball
\begin{equation}
\Psi\left(  r,\theta\right)  =\sum_{l=0}^{\infty}i^{l}~\frac{1}{2}B_{l}%
~j_{l}\left(  kr\right)  \left(  2l+1\right)  P_{l}\left(  \cos\theta\right)
\text{ \ \ for \ \ }0\leq R<r
\end{equation}
since the solution must be regular at $r=0$. \ The solutions on the cylinder
take the form%
\begin{equation}
\Psi\left(  h,\theta\right)  =\sum_{l=0}^{\infty}i^{l}~\frac{1}{2}\left(
C_{l}~\cos\left(  \kappa h\right)  +D_{l}~\sin\left(  \kappa h\right)
\right)  \left(  2l+1\right)  P_{l}\left(  \cos\theta\right)
\end{equation}
Continuity of $\Psi$ and its first derivatives on the edges of the cylinder
now determines all the coefficients. \ 

That is to say, the wave function must satisfy the boundary conditions%
\begin{align}
\left.  \Psi\right\vert _{h=0,\ r\searrow R}  &  =\left.  \Psi\right\vert
_{r=R,\ h\nearrow0}\text{ \ \ and \ \ }\left.  \frac{\partial}{\partial r}%
\Psi\right\vert _{h=0,\ r\searrow R}=\left.  \frac{\partial}{\partial h}%
\Psi\right\vert _{r=R,\ h\nearrow0}\\
\left.  \Psi\right\vert _{h=-H,\ r\nearrow R}  &  =\left.  \Psi\right\vert
_{r=R,\ h\searrow-H}\text{ \ \ and \ \ }\left.  \frac{\partial}{\partial
r}\Psi\right\vert _{h=-H,\ r\nearrow R}=\left.  \frac{\partial}{\partial
h}\Psi\right\vert _{r=R,\ h\searrow-H}%
\end{align}
Upon using the expansions for $\Psi$ on the three parts of the $n\rightarrow
\infty$ manifold, the coefficients are determined by these boundary
conditions. \ The results are%
\begin{align}
S_{l}\left(  x,y\right)   &  =-\frac{\operatorname*{numer}\left(
l,x,y\right)  }{\operatorname*{denom}\left(  l,x,y\right)  }\ ,\ \ \ B_{l}%
\left(  x,y\right)  =\frac{4i}{\pi}\frac{\sqrt{x^{2}-l\left(  l+1\right)  }%
}{\operatorname*{denom}\left(  l,x,y\right)  }\ ,\\
C_{l}\left(  x,y\right)   &  =+i\sqrt{\frac{8}{\pi x}}\frac{\sqrt
{x^{2}-l\left(  l+1\right)  }J_{l+1/2}\left(  x\right)  \cos\left(
2\sqrt{x^{2}-l\left(  l+1\right)  }y\right)  }{\operatorname*{denom}\left(
l,x,y\right)  }\ ,\\
D_{l}\left(  x,y\right)   &  =-i\sqrt{\frac{8}{\pi x}}\frac{\sqrt
{x^{2}-l\left(  l+1\right)  }J_{l+1/2}\left(  x\right)  \sin\left(
2y\sqrt{x^{2}-l\left(  l+1\right)  }\right)  }{\operatorname*{denom}\left(
l,x,y\right)  }\ ,
\end{align}
with the definitions (not to be confused with Cartesian coordinates for
$\mathbb{E}_{3}$!)
\begin{equation}
x\equiv kR\ ,\ \ \ y\equiv H/\left(  2R\right)  \ ,\ \ \ \kappa R=\sqrt
{x^{2}-l\left(  l+1\right)  }\equiv z\left(  l,x\right)  \ ,\ \ \ \kappa
H/2=y\sqrt{x^{2}-l\left(  l+1\right)  }=y~z\left(  l,x\right)
\end{equation}
and where the denominator and numerator functions are given by
\begin{align}
\operatorname*{denom}\left(  l,x,y\right)   &  =+\frac{2}{\pi}i\sqrt
{x^{2}-l\left(  l+1\right)  }\cos\left(  2y~z\left(  l,x\right)  \right) \\
&  +\left(  \left(  \left(  x^{2}-l\right)  J_{l+\frac{1}{2}}\left(  x\right)
-lxJ_{l+\frac{3}{2}}\left(  x\right)  \right)  H_{l+\frac{1}{2}}^{\left(
1\right)  }\left(  x\right)  +x\left(  xJ_{l+\frac{3}{2}}\left(  x\right)
-lJ_{l+\frac{1}{2}}\left(  x\right)  \right)  H_{l+\frac{3}{2}}^{\left(
1\right)  }\left(  x\right)  \right)  \sin\left(  2y~z\left(  l,x\right)
\right) \nonumber
\end{align}%
\begin{align}
\operatorname*{numer}\left(  l,x,y\right)   &  =-\frac{2}{\pi}i\sqrt
{x^{2}-l\left(  l+1\right)  }\cos\left(  2y~z\left(  l,x\right)  \right) \\
&  +\left(  \left(  \left(  x^{2}-l\right)  J_{l+\frac{1}{2}}\left(  x\right)
-lxJ_{l+\frac{3}{2}}\left(  x\right)  \right)  H_{l+\frac{1}{2}}^{\left(
2\right)  }\left(  x\right)  +x\left(  xJ_{l+\frac{3}{2}}\left(  x\right)
-lJ_{l+\frac{1}{2}}\left(  x\right)  \right)  H_{l+\frac{3}{2}}^{\left(
2\right)  }\left(  x\right)  \right)  \sin\left(  2y~z\left(  l,x\right)
\right) \nonumber
\end{align}
For real $x$ and real $y$ note that $\operatorname*{numer}\left(
l,x,y\right)  =\operatorname*{denom}^{\ast}\left(  l,x,y\right)  $ if
$x^{2}\geq l\left(  l+1\right)  $ and $\operatorname*{numer}\left(
l,x,y\right)  =-\operatorname*{denom}^{\ast}\left(  l,x,y\right)  $ if
$x^{2}\leq l\left(  l+1\right)  $. \ But either way, no matter whether
$x^{2}\geq l\left(  l+1\right)  $ or $x^{2}\leq l\left(  l+1\right)  $, the
scattering is elastic: $\left\vert S_{l}\left(  x,y\right)  \right\vert =1$.
\ The coefficients appearing in $f\left(  \theta\right)  $ and $\sigma_{el}$
are then%
\begin{align}
A_{l}  &  =\frac{i}{2}\left(  1-S_{l}\right) \\
&  =i\frac{\left(  \left(  \left(  x^{2}-l\right)  J_{l+\frac{1}{2}}\left(
x\right)  -lxJ_{l+\frac{3}{2}}\left(  x\right)  \right)  J_{l+\frac{1}{2}%
}\left(  x\right)  +x\left(  xJ_{l+\frac{3}{2}}\left(  x\right)
-lJ_{l+\frac{1}{2}}\left(  x\right)  \right)  J_{l+\frac{3}{2}}\left(
x\right)  \right)  \sin\left(  2y~z\left(  l,x\right)  \right)  }%
{\operatorname*{denom}\left(  l,x,y\right)  }\nonumber
\end{align}

For the $n\rightarrow\infty$ wormhole geometry, with incident flux only on the
\textquotedblleft upper\textquotedblright\ copy of $\mathbb{E}_{3}%
\smallsetminus\mathbb{B}_{3}\left(  R\right)  $, the difference with the
foxhole\ partial wave expansions is just the existence of purely outgoing wave
Bessel functions (i.e. $h_{l}^{\left(  1\right)  }\left(  kr\right)  $ but not
$h_{l}^{\left(  2\right)  }\left(  kr\right)  $) on the \textquotedblleft
lower\textquotedblright\ copy of $\mathbb{E}_{3}\smallsetminus\mathbb{B}%
_{3}\left(  R\right)  $ instead of the regular Bessel functions (i.e.
$j_{l}\left(  kr\right)  $) on the foxhole ball $\mathbb{B}_{3}\left(
R\right)  $. \ On that lower $\mathbb{E}_{3}\smallsetminus\mathbb{B}%
_{3}\left(  R\right)  $ the solution is now given by the expansion \
\begin{equation}
\Psi\left(  r,\theta\right)  =\sum_{l=0}^{\infty}i^{l}~\frac{1}{2}~T_{l}%
~h_{l}^{\left(  1\right)  }\left(  kr\right)  \left(  2l+1\right)
P_{l}\left(  \cos\theta\right)  \text{ \ \ for \ \ }R<r\leq\infty
\end{equation}
with coefficients $T_{l}$ to be determined along with $S_{l}$, $C_{l}$, and
$D_{l}$, where the latter are defined as in the foxhole expansions.
\ Corresponding expansions apply if there is incident flux only on the lower
3D space, of course, but only incident flux on the upper 3D space is
considered here.

The wave function must now satisfy the boundary conditions%
\begin{align}
\left.  \Psi\right\vert _{h=H,\ r\searrow R}  &  =\left.  \Psi\right\vert
_{r=R,\ h\nearrow H}\text{ \ \ and \ \ }\left.  \frac{\partial}{\partial
r}\Psi\right\vert _{h=H,\ r\searrow R}=\left.  \frac{\partial}{\partial h}%
\Psi\right\vert _{r=R,\ h\nearrow H}\\
\left.  \Psi\right\vert _{h=-H,\ r\searrow R}  &  =\left.  \Psi\right\vert
_{r=R,\ h\searrow-H}\text{ \ \ and \ \ }\left.  \frac{\partial}{\partial
r}\Psi\right\vert _{h=-H,\ r\searrow R}=-\left.  \frac{\partial}{\partial
h}\Psi\right\vert _{r=R,\ h\searrow-H}%
\end{align}
with a crucial minus sign in the last of these conditions. \ Upon using the
expansions for $\Psi$ on the three parts of the $n\rightarrow\infty$ manifold,
all the coefficients are determined by these boundary conditions. \ The
results most relevant for scattering are \
\begin{equation}
S_{l}\left(  x,y\right)  =-\frac{\operatorname*{wumer}\left(  l,x,y\right)
}{\operatorname*{wenom}\left(  l,x,y\right)  }\ ,\ \ \ T_{l}\left(
x,y\right)  =\frac{4i}{\pi}\frac{\sqrt{x^{2}-l\left(  l+1\right)  }%
}{\operatorname*{wenom}\left(  l,x,y\right)  }%
\end{equation}
where the wormhole denominator and numerator functions are given by%
\begin{align}
\operatorname*{wenom}\left(  l,x,y\right)   &  =2\sqrt{x^{2}-l\left(
l+1\right)  }H_{l+1/2}^{\left(  1\right)  }\left(  x\right)  \left(
lH_{l+1/2}^{\left(  1\right)  }\left(  x\right)  -xH_{l+3/2}^{\left(
1\right)  }\left(  x\right)  \right)  \cos\left(  2y~z\left(  l,x\right)
\right) \\
&  +\left(  \left(  x^{2}-l\left(  l+1\right)  \right)  \left(  H_{l+1/2}%
^{\left(  1\right)  }\left(  x\right)  \right)  ^{2}-\left(  lH_{l+1/2}%
^{\left(  1\right)  }\left(  x\right)  -xH_{l+3/2}^{\left(  1\right)  }\left(
x\right)  \right)  ^{2}\right)  \sin\left(  2y~z\left(  l,x\right)  \right)
\nonumber
\end{align}%
\begin{gather}
\operatorname*{wumer}\left(  l,x,y\right)  =\\
\sqrt{x^{2}-l\left(  l+1\right)  }\left(  H_{l+1/2}^{\left(  2\right)
}\left(  x\right)  \left(  lH_{l+1/2}^{\left(  1\right)  }\left(  x\right)
-xH_{l+3/2}^{\left(  1\right)  }\left(  x\right)  \right)  +H_{l+1/2}^{\left(
1\right)  }\left(  x\right)  \left(  lH_{l+1/2}^{\left(  2\right)  }\left(
x\right)  -xH_{l+3/2}^{\left(  2\right)  }\left(  x\right)  \right)  \right)
\cos\left(  2y~z\left(  l,x\right)  \right) \nonumber\\
+\left(  \left(  x^{2}-l\left(  l+1\right)  \right)  H_{l+1/2}^{\left(
2\right)  }\left(  x\right)  H_{l+1/2}^{\left(  1\right)  }\left(  x\right)
-\left(  lH_{l+1/2}^{\left(  2\right)  }\left(  x\right)  -xH_{l+3/2}^{\left(
2\right)  }\left(  x\right)  \right)  \left(  lH_{l+1/2}^{\left(  1\right)
}\left(  x\right)  -xH_{l+3/2}^{\left(  1\right)  }\left(  x\right)  \right)
\right)  \sin\left(  2y~z\left(  l,x\right)  \right) \nonumber
\end{gather}
with the definitions (again, not to be confused with Cartesian coordinates for
$\mathbb{E}_{3}$!)%
\begin{equation}
x\equiv kR\ ,\ \ \ y\equiv H/R\ ,\ \ \ \kappa R=\sqrt{x^{2}-l\left(
l+1\right)  }\equiv z\left(  l,x\right)  \ ,\ \ \ \kappa H=y\sqrt
{x^{2}-l\left(  l+1\right)  }=y~z\left(  l,x\right)
\end{equation}
Note the definition of the aspect ratio $y$ in terms of $H$ and $R$ for the
wormhole differs from that for the foxhole by a factor of $2$.

For the wormhole with incident waves only on the upper 3D space, it does not
matter if $k^{2}R^{2}\geq l\left(  l+1\right)  $ so that $\kappa R=\sqrt
{k^{2}R^{2}-l\left(  l+1\right)  }$ is real, or if $k^{2}R^{2}\leq l\left(
l+1\right)  $ so that $\kappa R=i\sqrt{l\left(  l+1\right)  -k^{2}R^{2}}$ is
imaginary, either way $\left\vert S_{l}\right\vert \neq1$. \ This means there
is not only elastic scattering on the upper 3D space but also absorption by
the wormhole with outward flow on the bottom 3D space. \ The elastic cross
section for scattered waves on the upper 3D space is given by%
\begin{equation}
\sigma_{el}=\frac{\pi}{k^{2}}\sum_{l=0}^{\infty}\left(  2l+1\right)
\left\vert S_{l}-1\right\vert ^{2}=\frac{\pi}{k^{2}}\sum_{l=0}^{\infty}\left(
2l+1\right)  \left(  1+\left\vert S_{l}\right\vert ^{2}-2\operatorname{Re}%
S_{l}\right)
\end{equation}
while the inelastic (absorption) and total cross-sections are given by%
\begin{equation}
\sigma_{inel}=\frac{\pi}{k^{2}}\sum_{l=0}^{\infty}\left(  2l+1\right)  \left(
1-\left\vert S_{l}\right\vert ^{2}\right)  \ ,\ \ \ \sigma_{tot}=\sigma
_{el}+\sigma_{inel}=\frac{2\pi}{k^{2}}\sum_{l=0}^{\infty}\left(  2l+1\right)
\left(  1-\operatorname{Re}S_{l}\right)
\end{equation}
Since the model is invariant under $\left.  \mathbb{E}_{3}\right\vert
_{\text{upper}}\leftrightarrow\left.  \mathbb{E}_{3}\right\vert _{\text{lower}%
}$, if there is incident flux only on the lower 3D space, with corresponding
expansion coefficients, the same elastic and inelastic cross-sections are obtained.

\section{Resonance poles}

The foxhole scattering amplitudes exhibit poles in the lower half complex $k$
(or $x\equiv kR$) plane, in complete accord with general theory for potential
scattering \cite{Newton}. For the $n\rightarrow\infty$ foxhole geometry these
poles can be obtained to arbitrary accuracy by numerical computation of the
$\operatorname*{denom}\left(  l,x,y\right)  $ zeroes. Alternatively, a simple
estimate of the corresponding resonance frequencies follows from the
Bohr-Sommerfeld (B-S) method \cite{BohrSommerfeld}. \ 

The cylinder is all that distinguishes the $n\rightarrow\infty$ foxhole
geometry from flat space. \ On the cylinder the classical action for a
complete \textquotedblleft up and down\textquotedblright\ cycle of motion
along the cylinder length is given by a trivial integral, since momentum is
constant on the cylinder.%
\begin{equation}
I=2\int_{-H}^{0}\kappa dh=2H\sqrt{k^{2}-\frac{l\left(  l+1\right)  }{R^{2}}%
}=4y\sqrt{x^{2}-l\left(  l+1\right)  }%
\end{equation}
with $x=kR$ and $y=H/\left(  2R\right)  $. \ Quantizing this action as
$I=\left(  2n+1\right)  \pi$ for $n=0,1,2,\cdots$ then gives an estimate of
the various resonant $x$ values as
\begin{equation}
x_{res}=\sqrt{l\left(  l+1\right)  +\frac{\left(  2n+1\right)  ^{2}\pi^{2}%
}{16y^{2}}}%
\end{equation}
For $l=0$ this estimate gives $x_{\text{nth }l=0\text{ resonance}}=\left(
2n+1\right)  \pi/\left(  4y\right)  $. \ For example, consider $y=2$ for
$n=0,\ 1,\ 2,\ \cdots$, to find
\begin{equation}
x_{l=0\text{ resonance}}=\pi/8\ ,\ \ \ 3\pi/8\ ,\ \ \ 5\pi/8,\ \ \ 7\pi
/8\ ,\ \ \ \cdots\text{ \ \ }%
=0.3927\ ,\ \ \ 1.178\ ,\ \ \ 1.963\ ,\ \ \ 2.749\ ,\ \ \ \cdots
\end{equation}
The accuracy of these estimates for $l=0$ leaves something to be desired, as
evident in the numerical solution of $0=\operatorname*{denom}\left(
0,x,2\right)  $ as well as numerical plots of the partial cross-section
$\sigma_{0}\left(  x,2\right)  $.

However, the B-S estimate for the cross-section peaks improves as $kR$ becomes
large. \ To understand this, consider the principal asymptotic forms for the
Bessel functions that make up the scattering amplitudes. \
\begin{align}
&  J_{l+\frac{1}{2}}\left(  x\right)  \underset{x\gg l}{\sim}\sqrt{\frac
{2}{\pi x}}\sin\left(  x-\frac{\pi}{2}l\right)  \ ,\ \ \ Y_{l+\frac{1}{2}%
}\left(  x\right)  \underset{x\gg l}{\sim}-\sqrt{\frac{2}{\pi x}}\cos\left(
x-\frac{\pi}{2}l\right) \\
&  H_{l+\frac{1}{2}}^{\left(  1\right)  }\left(  x\right)  \underset{x\gg
l}{\sim}-i\sqrt{\frac{2}{\pi x}}e^{i\left(  x-\frac{\pi}{2}l\right)
}\ ,\ \ \ H_{l+\frac{1}{2}}^{\left(  2\right)  }\left(  x\right)
\underset{x\gg l}{\sim}i\sqrt{\frac{2}{\pi x}}e^{-i\left(  x-\frac{\pi}%
{2}l\right)  }%
\end{align}
It follows that the leading asymptotic behavior of $\operatorname*{denom}%
\left(  l,x,y\right)  $ is%
\begin{equation}
\operatorname*{denom}\left(  l,x,y\right)  \underset{x\gg l}{\sim}\frac
{2ix}{\pi}\exp\left(  -2iy\sqrt{x^{2}-l\left(  l+1\right)  }\right)
\end{equation}
with real and imaginary parts%
\begin{align}
&  \operatorname{Re}\left(  \operatorname*{denom}\left(  l,x,y\right)
\right)  \underset{x\gg l}{\sim}\frac{2x}{\pi}~\sin\left(  2y\sqrt
{x^{2}-l\left(  l+1\right)  }\right) \\
&  \operatorname{Im}\left(  \operatorname*{denom}\left(  l,x,y\right)
\right)  \underset{x\gg l}{\sim}\frac{2x}{\pi}~\cos\left(  2y\sqrt
{x^{2}-l\left(  l+1\right)  }\right)
\end{align}
Roots of the asymptotic form for $\operatorname{Im}\left(
\operatorname*{denom}\left(  l,x,y\right)  \right)  $ then yield the B-S
quantization condition: \ $2y\sqrt{x^{2}-l\left(  l+1\right)  }=\left(
n+1/2\right)  \pi$.

On the other hand, evaluating the asymptotic form of $\operatorname{Re}\left(
\operatorname*{denom}\left(  l,x,y\right)  \right)  $ at these roots then
gives
\begin{equation}
\left.  \operatorname{Re}\left(  \operatorname*{denom}\left(  l,x,y\right)
\right)  \right\vert _{\text{B-S}}\underset{x\gg l}{\sim}\frac{2x}{\pi}\left(
-1\right)  ^{n}%
\end{equation}
which does \emph{not} vanish. \ But in fact, this is just what should have
been expected, since the asymptotic form of the scattering amplitude itself is%
\begin{equation}
S_{l}\left(  x,y\right)  \underset{x\gg l}{\sim}\exp\left(  4iy\sqrt
{x^{2}-l\left(  l+1\right)  }\right)  =\exp\left(  2iH\sqrt{k^{2}%
-\frac{l\left(  l+1\right)  }{R^{2}}}\right)
\end{equation}
where once again $y=\frac{H}{2R}$. \ This is just the additional phase
acquired by a plane wave with momentum $\kappa=\sqrt{k^{2}-\frac{l\left(
l+1\right)  }{R^{2}}}$ as it goes down the cylinder length, and then back up
the cylinder length, with no other net phase change from its encounters with
both the upper and lower edges of the cylinder. \ 

Evaluating at the B-S condition gives%
\begin{equation}
\left.  S_{l}\left(  x,y\right)  \right\vert _{\text{B-S}}\underset{x\gg
l}{\sim}\exp\left(  \left(  2n+1\right)  i\pi\right)  =-1
\end{equation}
This value for $S_{l}$ maximizes $\left\vert A_{l}\right\vert $ and the
partial cross-section. \ To be more explicit, the asymptotic form of the $l$th
partial wave total cross-section for scattering from the foxhole is given by
\begin{equation}
\frac{\sigma_{l}\left(  x,y\right)  }{\pi R^{2}}\underset{x\gg l}{\sim}%
\frac{2}{x^{2}}\left(  2l+1\right)  \left(  1-\operatorname{Re}\left(
\exp\left(  4iy\sqrt{x^{2}-l\left(  l+1\right)  }\right)  \right)  \right)
=\frac{4}{x^{2}}\left(  2l+1\right)  \sin^{2}\left(  2y\sqrt{x^{2}-l\left(
l+1\right)  }\right)
\end{equation}
Evaluating at the B-S condition then saturates the unitarity bound for this
partial cross-section:%
\begin{equation}
\left.  \frac{\sigma_{l}\left(  x,y\right)  }{\pi R^{2}}\right\vert
_{\text{B-S}}\underset{x\gg l}{\sim}\frac{4}{x^{2}}\left(  2l+1\right)
\end{equation}
The asymptotic partial cross-section is therefore maximized at the B-S value
for $kR$.

\newpage

\section{Cross-sections}

Consider the $n\rightarrow\infty$ foxhole with aspect ratio $y=2$, and compute
numerically the net elastic cross-section for $x\leq6$. \ This is given
accurately by $\sigma_{el}=\sum_{l=0}^{5}\sigma_{l}\left(  x,2\right)  $ since
for this aspect ratio the higher $l$ partial waves only contribute
significantly for $x>6$. \ The numerical results are as shown here.

\noindent%

\begin{figure}[ht]
	\centering
	\begin{minipage}[t]{0.32\textwidth}
		\centering
		\includegraphics[width=\linewidth,height=6cm]{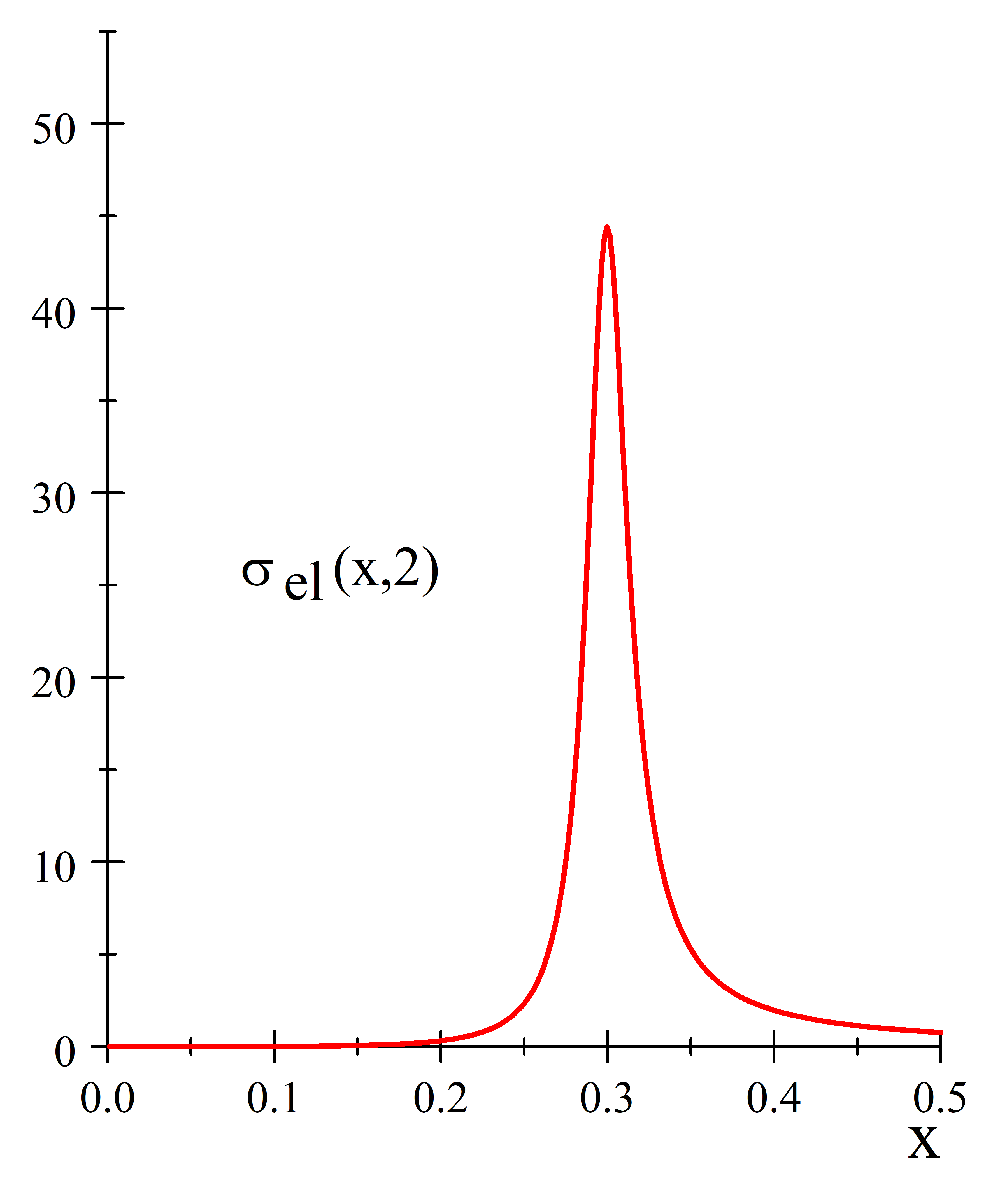}
		\caption{Fox Hole Scattering}
	\end{minipage}
	\hfill
	\begin{minipage}[t]{0.66\textwidth}
		\centering
		\includegraphics[width=\linewidth,height=6cm]{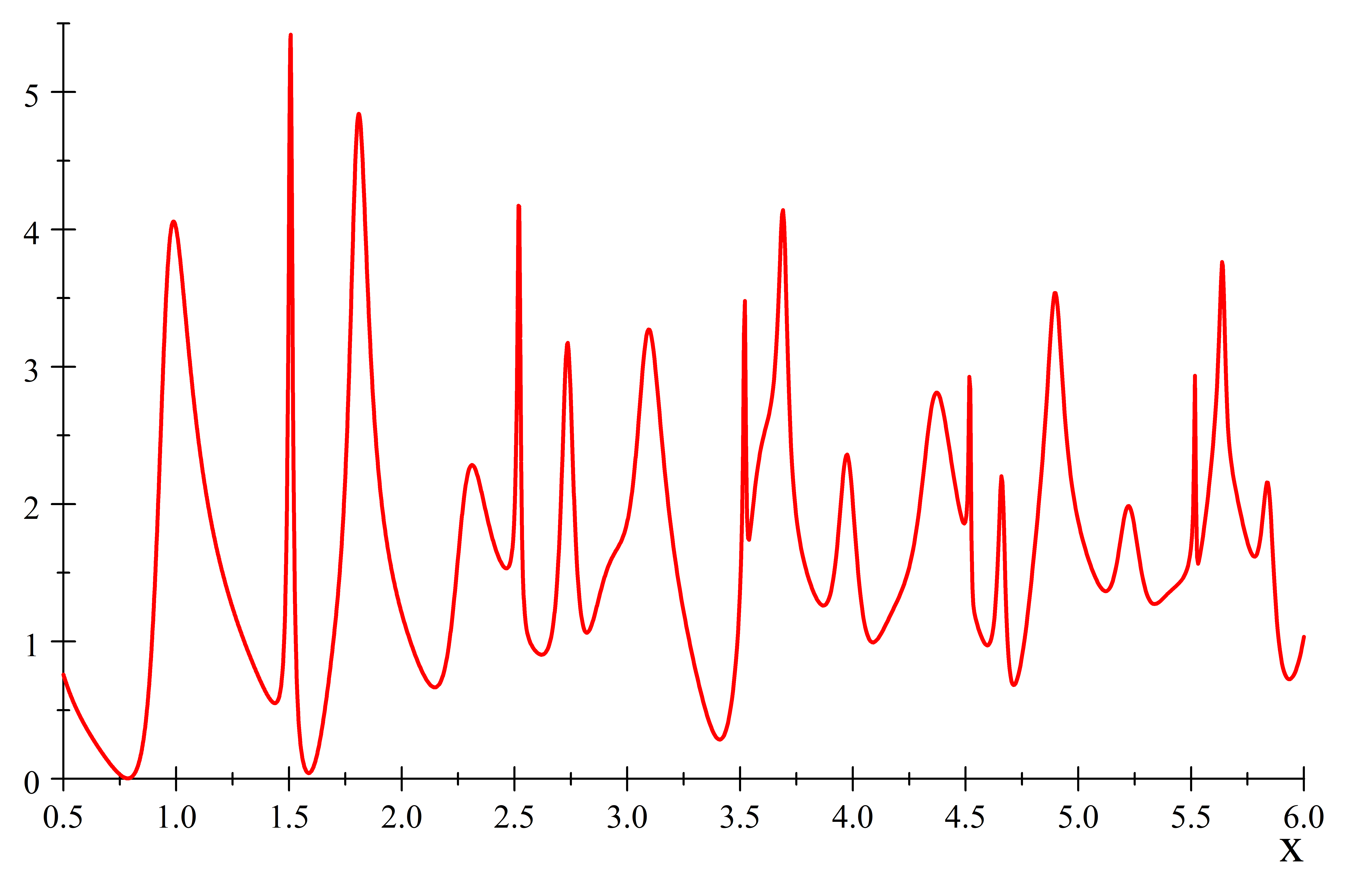}
		\caption{Fox Hole Scattering}
	\end{minipage}
\end{figure}

\noindent Note the change in vertical scales in these two graphs. \ Also
recall that $\sigma_{tot}=\sigma_{el}$ for real aspect ratios.

As happens in the case of Mie scattering for real index of refraction, the
complicated structure exhibited by the foxhole total cross-section can be
better understood by considering the individual partial wave contributions.
\ The first few of these partial cross-sections are plotted below. \ The light
gray curves are unitarity bounds, and the light blue lines indicate B-S
approximations for the real parts of the resonance poles\ located in the lower
half of the complex $k$ plane, as discussed in the previous Section. \ For
example, consider $l=0$.

\noindent%

\begin{figure}[ht]
	\centering
	\begin{minipage}[t]{0.32\textwidth}
		\centering
		\includegraphics[width=\linewidth,height=6cm]{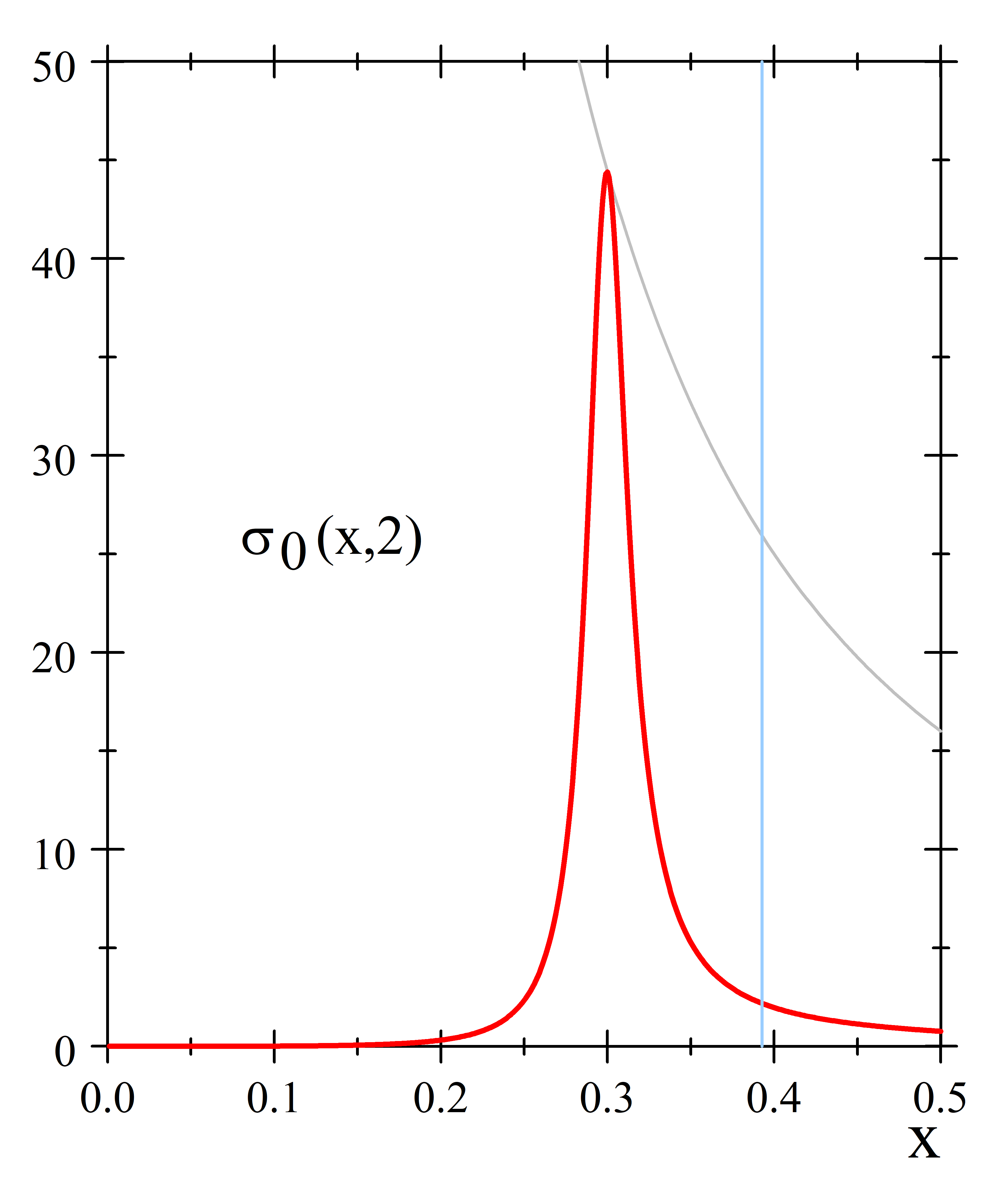}
		\caption{Resonance for x$\leq$0.5 }
	\end{minipage}
	\hfill
	\begin{minipage}[t]{0.66\textwidth}
		\centering
		\includegraphics[width=\linewidth,height=6cm]{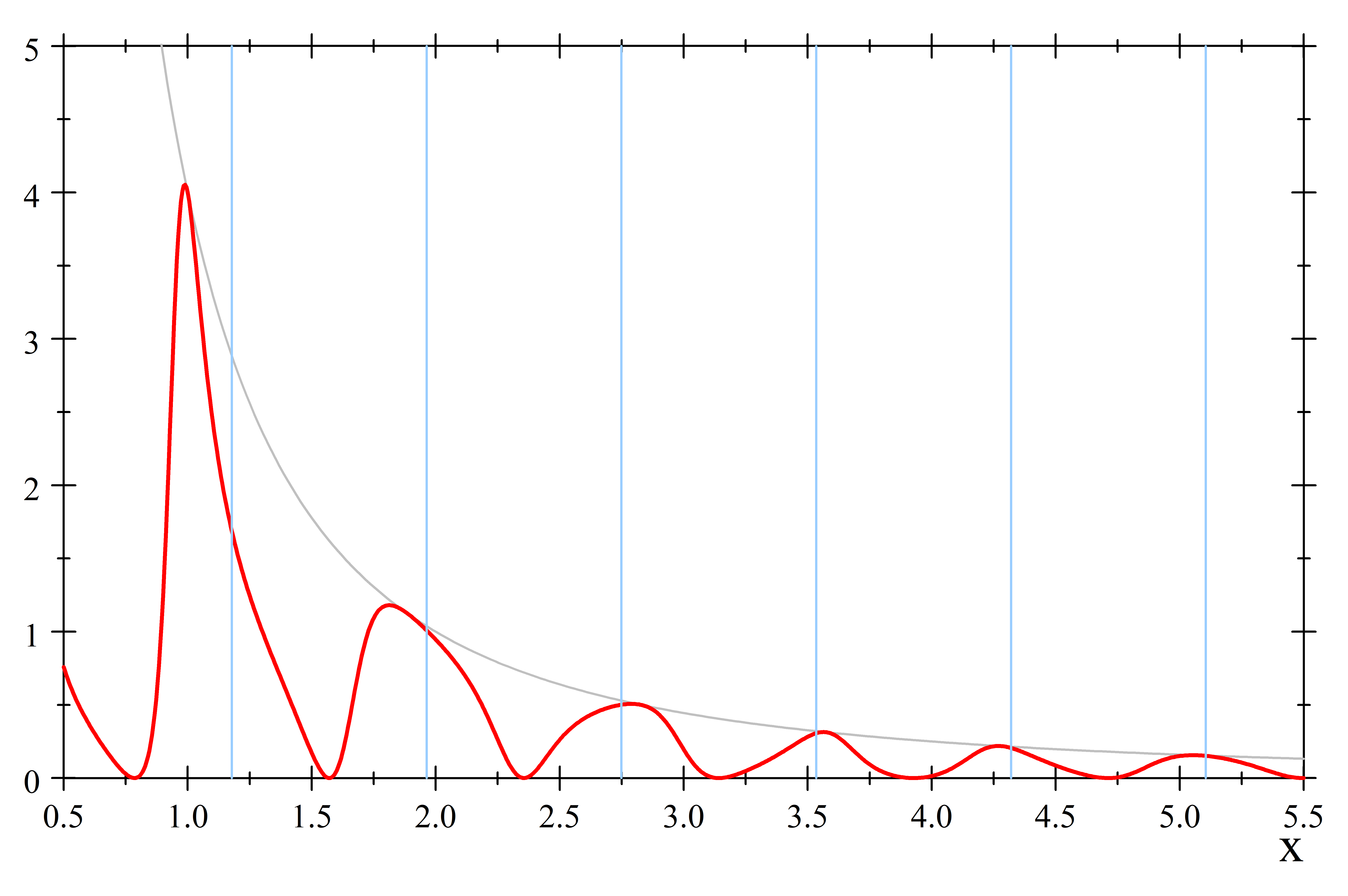}
		\caption{Resonances for 0.5$\leq$x$\leq$6}
	\end{minipage}
\end{figure}

\noindent Again, note the change in vertical scales in these two graphs. \ The
B-S approximation does not give accurate locations for the lowest two or three
$l=0$ peaks, although it is more accurate for the higher $l=0$ peaks as well
as for $l>0$ resonances, as evident in the graphs to follow. \ Nonetheless, in
all cases the B-S approximation does predict the correct \emph{number} of
peaks. \ Consider $\sigma_{l}\left(  x,2\right)  $ for $l=1,\ 2,\ $\& $3$.
\newpage

\begin{figure}[H]
	\centering
	\begin{minipage}[c]{0.75\textwidth}
		\centering
		\includegraphics[height=7cm]{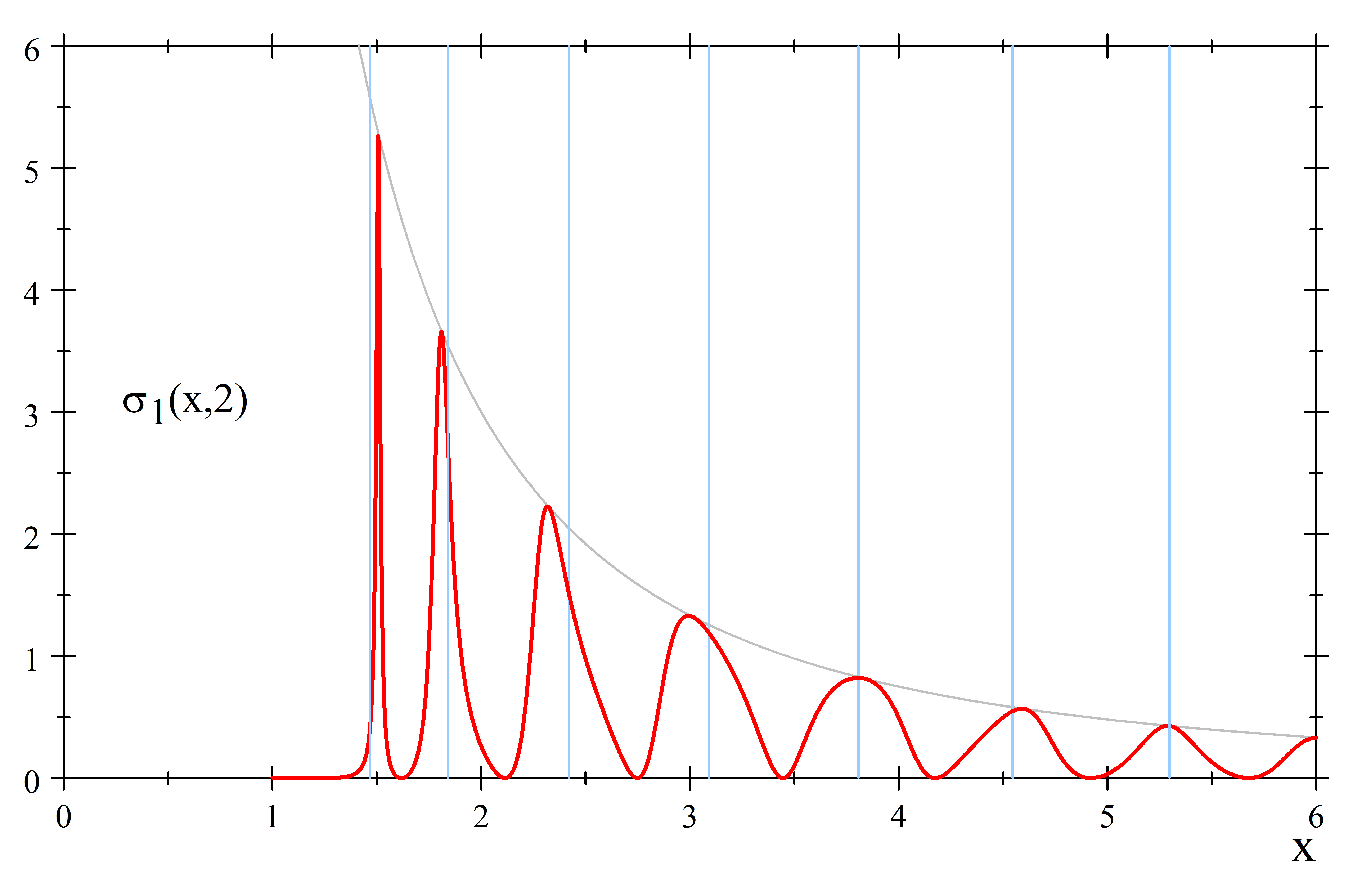}
		\caption{for l=1}
	\end{minipage}%
	\vspace{5mm}
	\begin{minipage}[c]{0.75\textwidth}
		\centering
		\includegraphics[height=7cm]{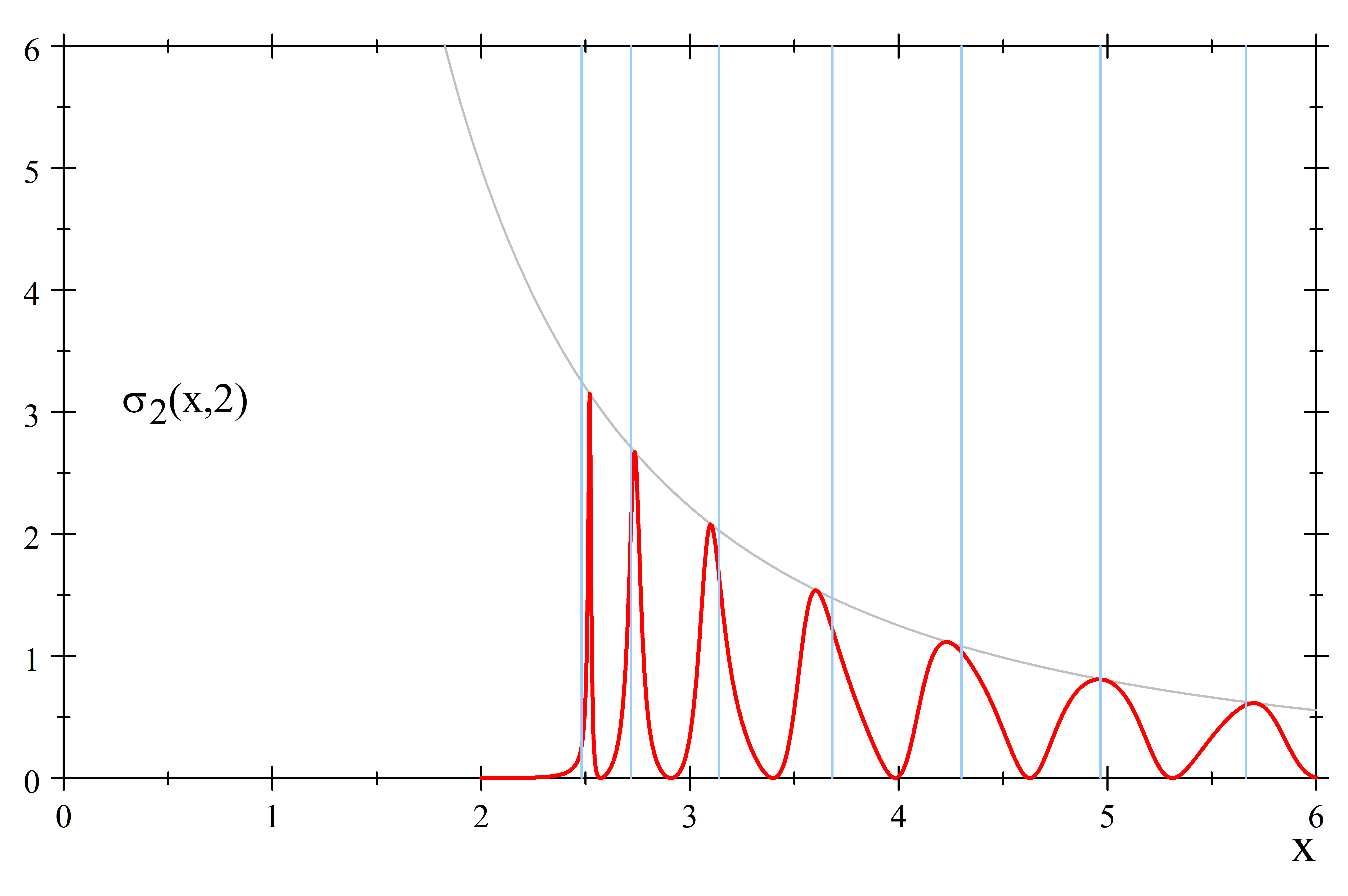}
		\caption{for l=2}
	\end{minipage}
	\vspace{5mm}
\begin{minipage}[c]{0.75\textwidth}
	\centering
	\includegraphics[height=7cm]{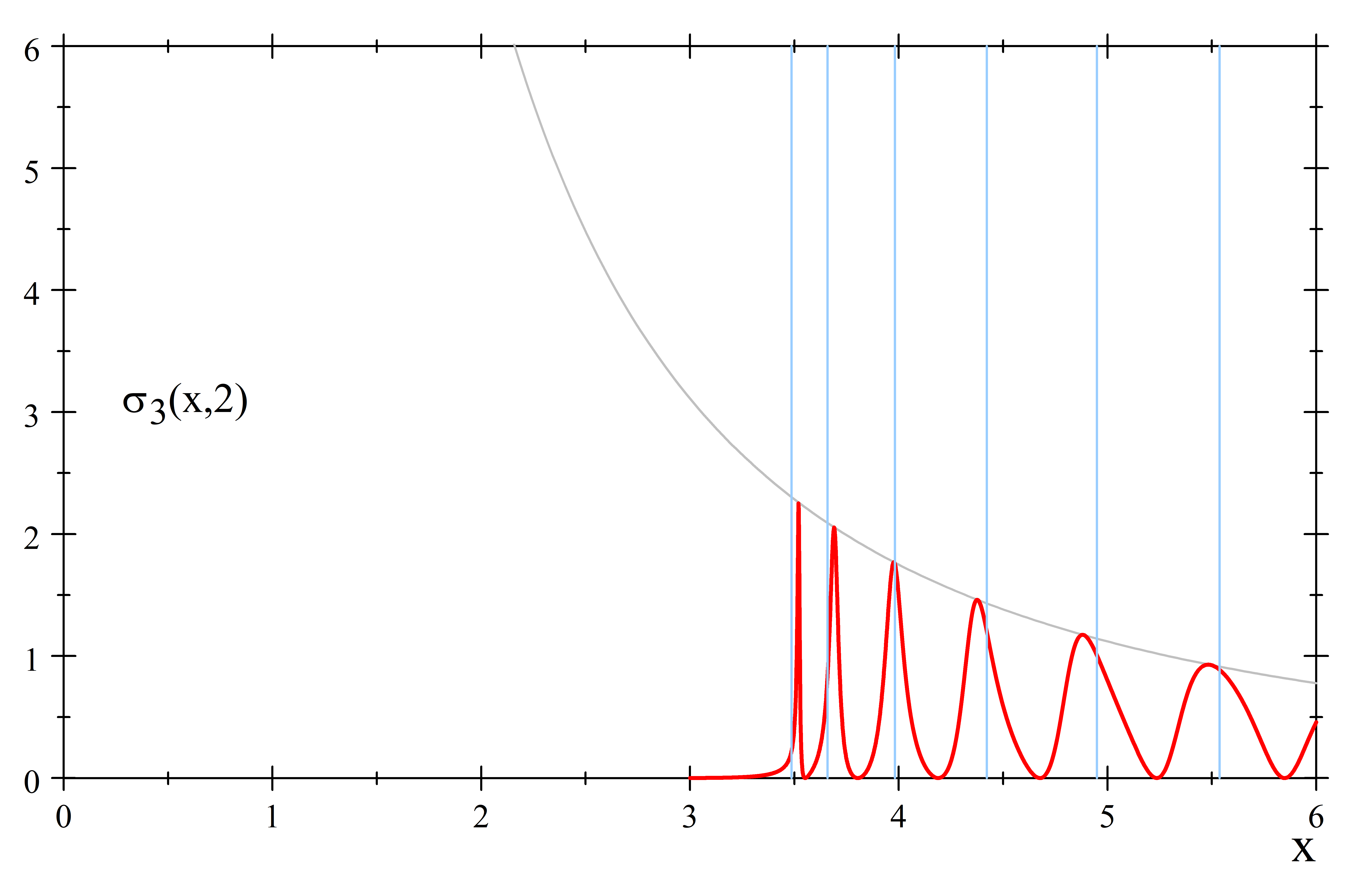}
	\caption{for l=3}
\end{minipage}%
\end{figure}

\newpage
The accuracy of the approximation $\sigma_{el}=\sum_{l=0}^{5}\sigma_{l}\left(
x,2\right)  $ for $x\leq6$ is evident in the following graph of $\sigma
_{6}\left(  x,2\right)  $.%
\begin{figure}[H]
	\centering
	\includegraphics[width=4.7in]{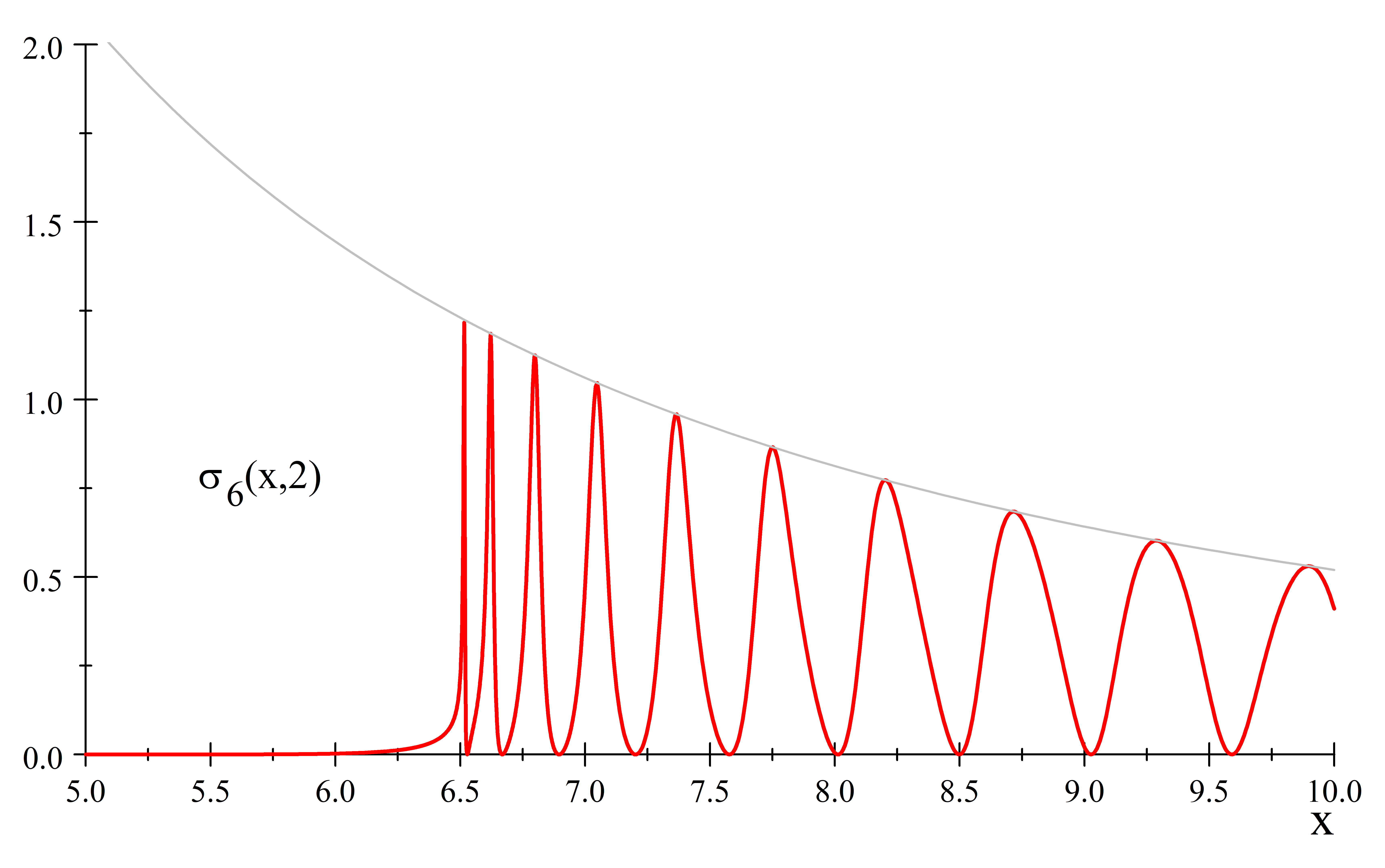} 
	\caption{for l=6}
\end{figure}

Numerical results for the infinite $n$ wormhole geometry defined above are
easily obtained from the exact partial wave expansions of the previous
sections, evincing qualitative features similar to those of the foxhole
example, but with the added feature of non-zero inelastic scattering (some
flux goes \textquotedblleft down the drain\textquotedblright) \cite{S,CSunpub}%
. \ 

\section{Tractable Generalizations}

There are several generalizations of the geometries given here that warrant
further study. \ For either the foxhole or wormhole models, a tractable,
easily solvable modification of the infinite $n$ geometry in Section 3\ would
be to have different radii for the spheres at the top and bottom of the
cylindrical section, i.e. the corresponding equatorial slice of the cylinder
would be a trapezoid. \ Such modifcations would perhaps convey intuition about
related dielectric systems. \ If orthogonally projected onto extensions of the
ambient Euclidean space (say, as viewed \textquotedblleft from
above\textquotedblright\ in diagrams similar to those in Figures 1 and 2), the
resulting projection of wavelets on the trapezoidal cylinder would appear to
have shorter wavelengths, hence\ lower phase velocities, thereby corresponding
to effective refraction indices with magnitude greater than unity. \ Moreover,
if the lower radius is greater than the upper radius, the effective projected
phase velocity of wavelets on the cylindrical portion of the trapezoidal
foxhole would appear to be negative. \ 

\section{Summary}

Scattering due to nontrivial spherically symmetric spatial geometries, and due
to nothing else, has been illustrated in this paper for a selection of simple
geometries. \ The analysis was non-relativistic in the sense that it was
framed in the mathematical context of the three-dimensional Helmholtz equation
where only spatial geometry was modified from Euclidean space, with\ time
taken to be universal. The emphasis here was on the geometry and the resulting
cross-sections as computed directly from exact results for partial wave expansions.

Relevant formulae were given in the body of the paper along with selected
numerical graphs to illustrate various important physical effects. \ Among
those effects were the occurrence of resonances and the saturation of
unitarity bounds.

Additional plots for comparison to various well-known situations have been
collected in an Appendix.\ \bigskip
\newpage
\noindent\textbf{Acknowledgements} TC received financial support from the United States Social Security Administration.

\noindent\textbf{Disclosures} The authors declare no conflicts of interest.

\appendix
\renewcommand{\thefigure}{A-\arabic{figure}}
\setcounter{figure}{0}

\section{Appendix: \ Perfect spheres}%
\begin{figure}[H]
	\centering
	\begin{minipage}[c]{0.75\textwidth}
		\centering
		\includegraphics[height=8cm]{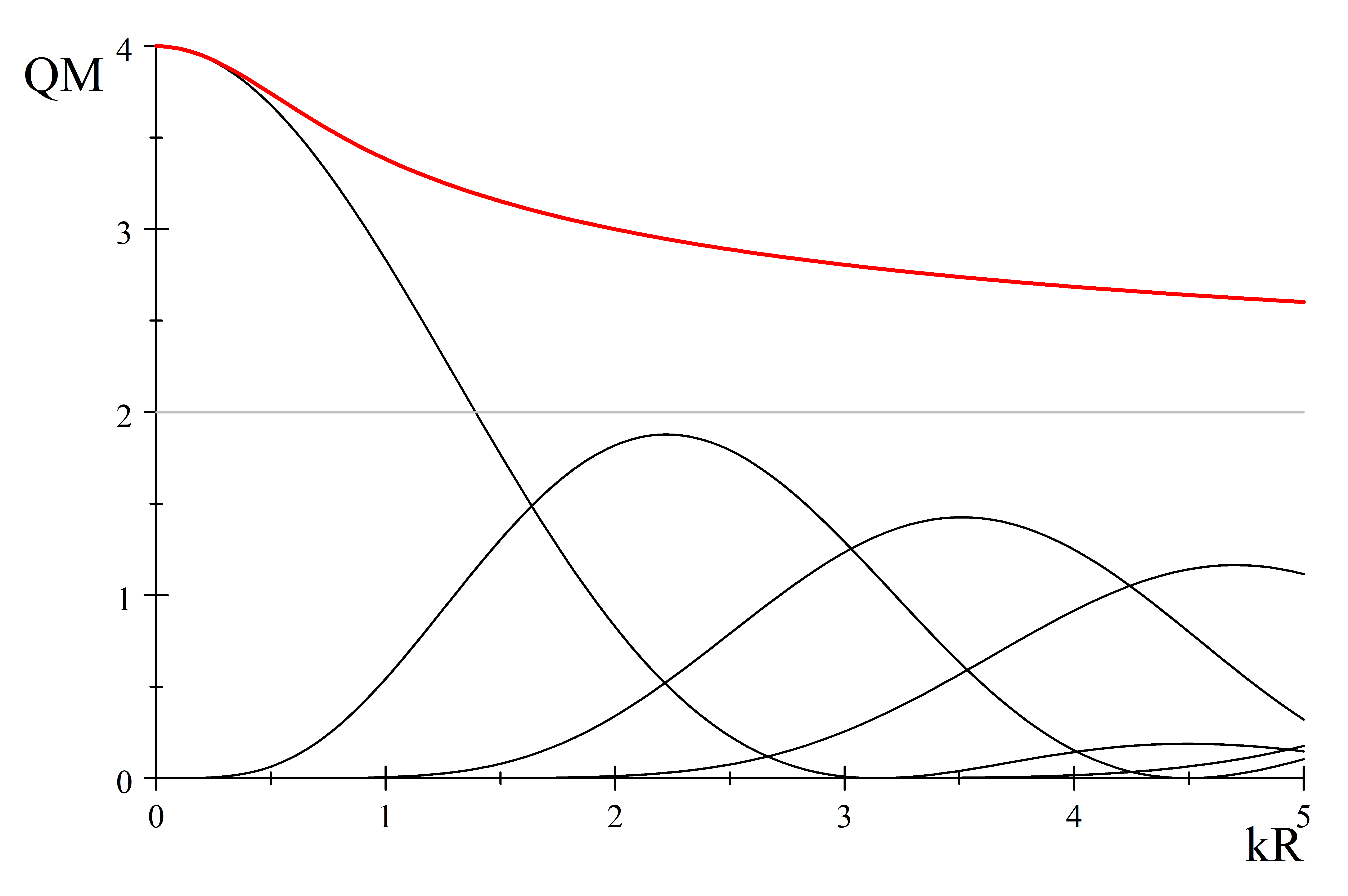}
\caption{
	In black, individual $\sigma_{l}/\left( \pi R^{2} \right)$ for $l=0,1,2$ \& $3$ from left to right.\\
	In red, $\sigma_{\text{QM hard sphere}}/\left( \pi R^{2} \right) = \sum_{l=0}^{\infty} \sigma_{l\text{ QM hard sphere}}/\left( \pi R^{2} \right)$\\
	but only $l\leq5$ contribute significantly for $kR\leq5$.
}
	\end{minipage}%
	\vspace{5mm}
	\begin{minipage}[c]{0.75\textwidth}
		\centering
		\includegraphics[height=8cm]{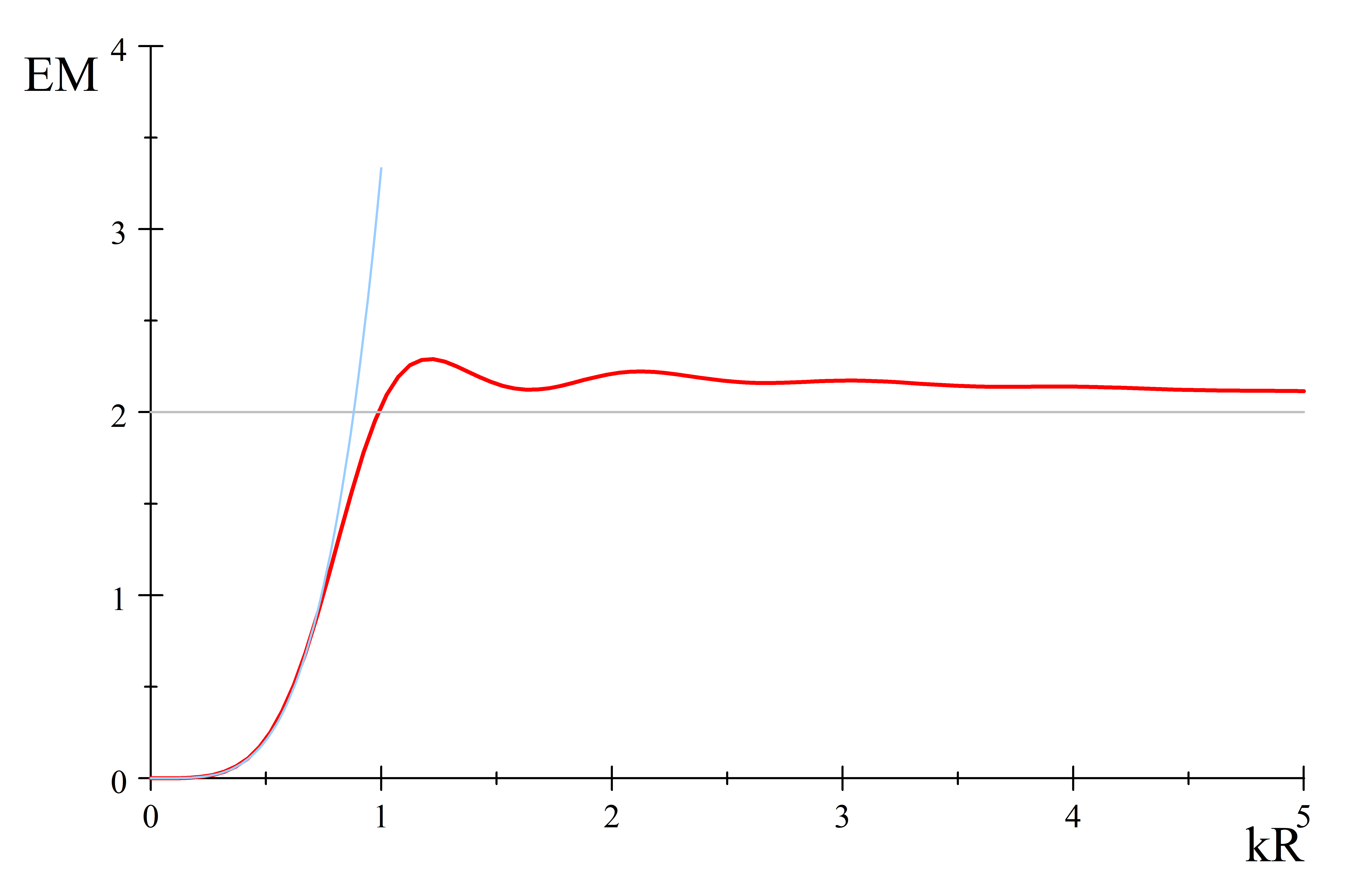}
		\caption{$\frac{1}{\pi R^{2}}\left(  \sigma_{\text{TE perf cond sphere}}+\sigma
			_{\text{TM perf cond sphere}}\right)  $ and\\ $\frac{1}{\pi
				R^{2}}~\sigma_{\text{quasi-static}}=\frac{10}{3}k^{4}R^{4}$ (light blue).}
	\end{minipage}
\end{figure}

\newpage

\begin{figure}[H]
	\centering
	\begin{minipage}[c]{0.75\textwidth}
		\centering
		\includegraphics[height=7cm]{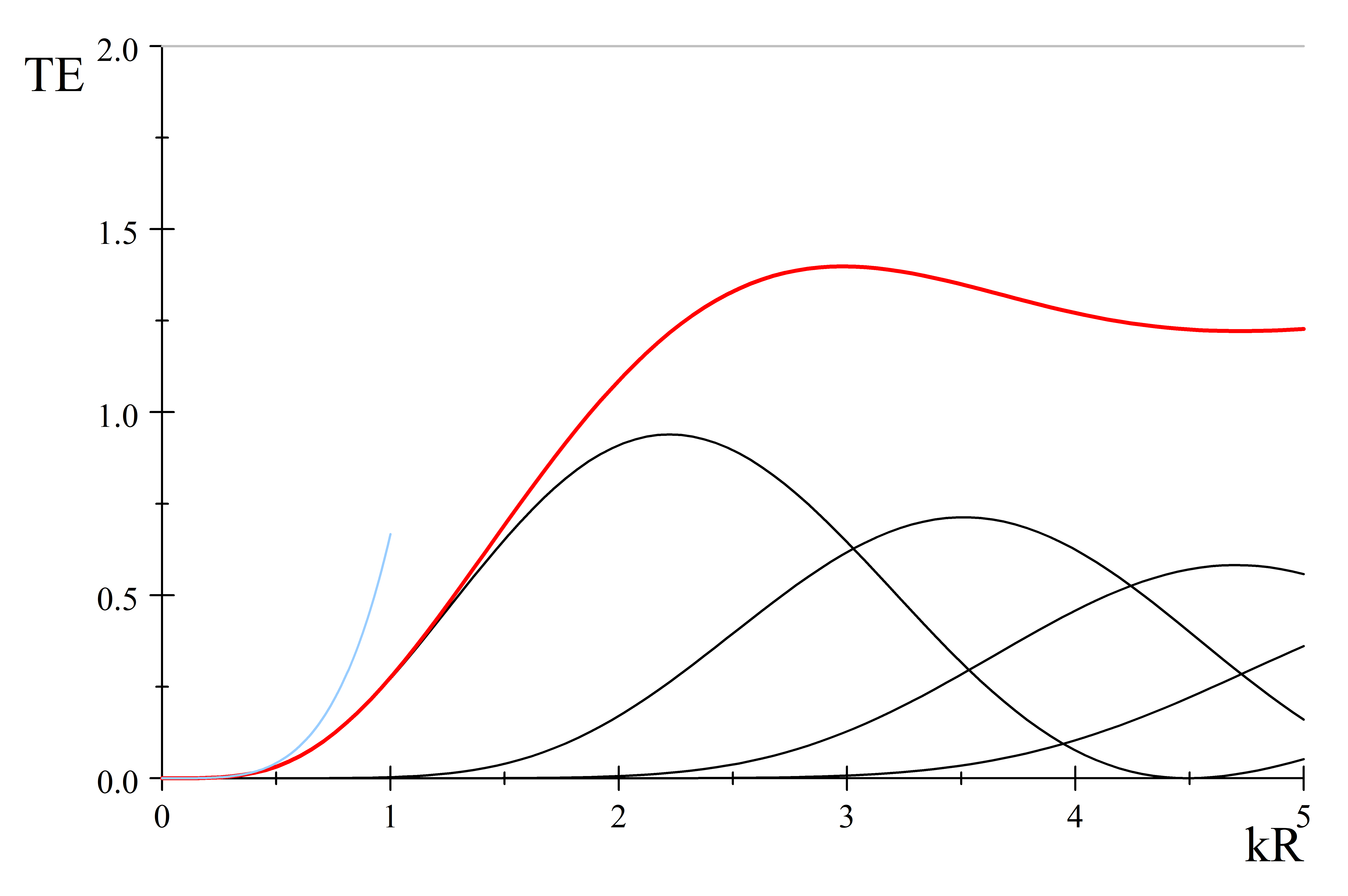}
		\caption{$\frac{1}{\pi R^{2}}~\sigma_{\text{TE perf cond sphere}}$ and $\frac{2}%
			{3}k^{4}R^{4}$ (light blue)}
	\end{minipage}%
	\vspace{5mm}
	\begin{minipage}[c]{0.75\textwidth}
		\centering
		\includegraphics[height=7cm]{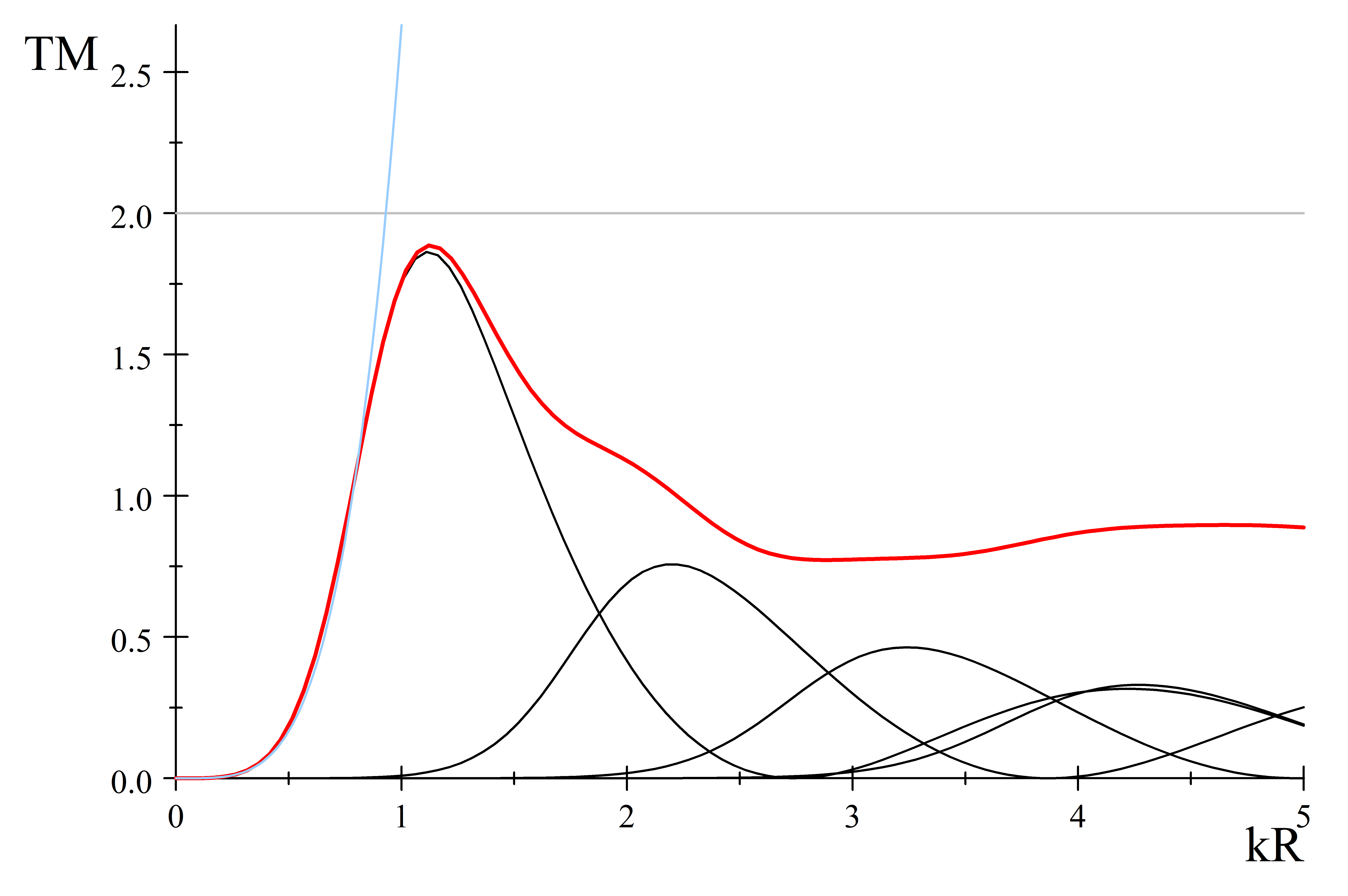}
		\caption{$\frac{1}{\pi R^{2}}~\sigma_{\text{TE perf cond sphere}}$ and $\frac{2}%
			{3}k^{4}R^{4}$ (light blue)}
	\end{minipage}
\end{figure}

\newpage

For dielectric spheres, consider the case $n=2$ for numerical purposes.%

\begin{figure}[H]
	\centering
	\begin{minipage}[c]{0.75\textwidth}
		\centering
		\includegraphics[height=7cm]{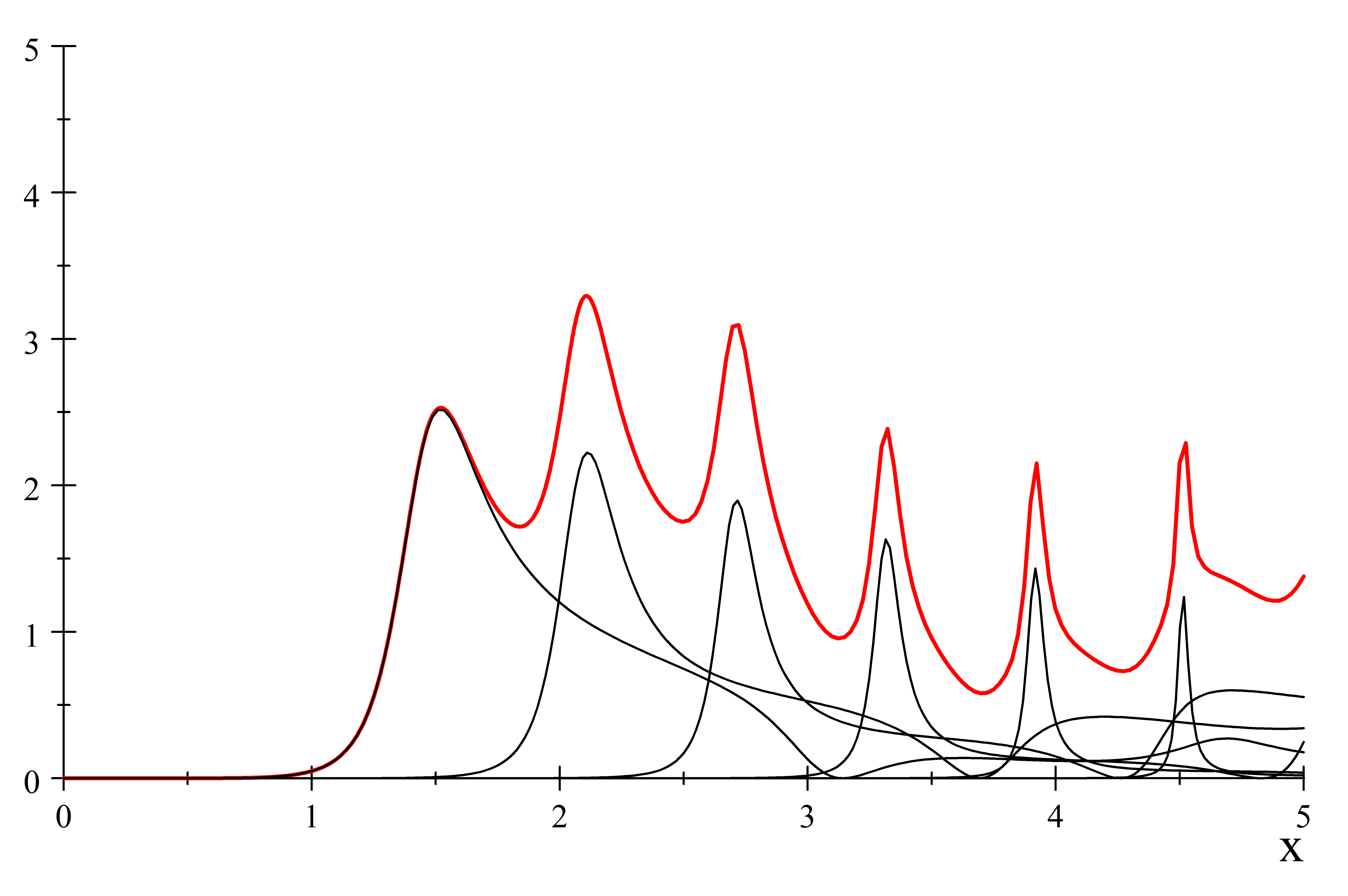}
		\caption{$l=1$ to $6$ summed contributions to $\sigma_{TE}$}
	\end{minipage}%
	\vspace{5mm}
	\begin{minipage}[c]{0.75\textwidth}
		\centering
		\includegraphics[height=7cm]{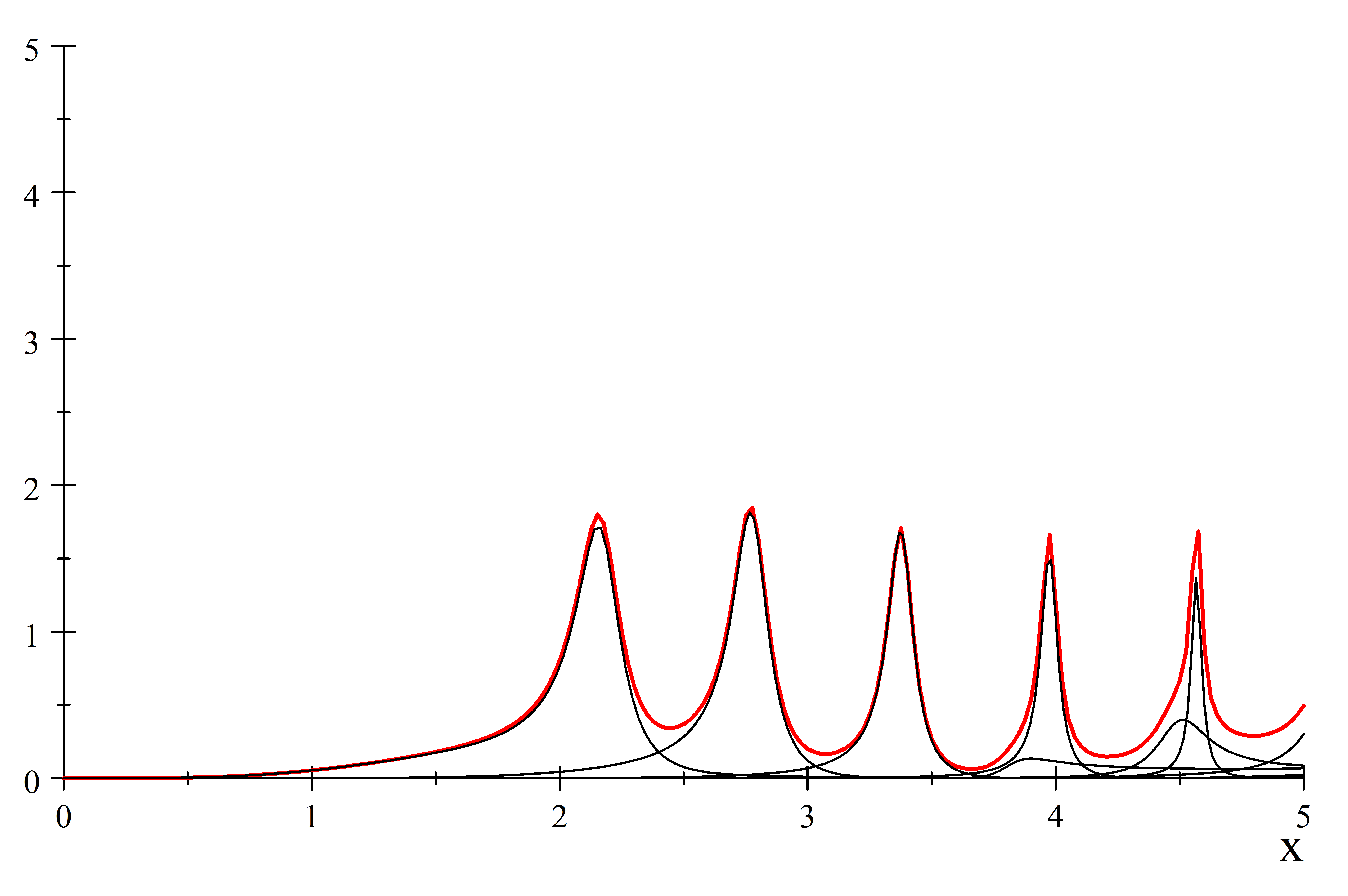}
		\caption{$l=1$ to $6$ summed contributions to $\sigma_{TM}$.}
	\end{minipage}
\end{figure}

Also, for real $n$, the so-called \textquotedblleft physical
optics\textquotedblright\ approximation for Mie scattering is \cite{Zangwill}%
\
\[
\sigma_{P}\left(  x,n\right)  =2\left(  1-\frac{2}{2x\left(  n-1\right)  }%
\sin\left(  2x\left(  n-1\right)  \right)  +\frac{2}{\left(  2x\left(
n-1\right)  \right)  ^{2}}\left(  1-\cos\left(  2x\left(  n-1\right)  \right)
\right)  \right)
\]%
\newpage
\begin{figure}[H]
	\centering
	\begin{minipage}[c]{0.75\textwidth}
		\centering
		\includegraphics[height=7cm]{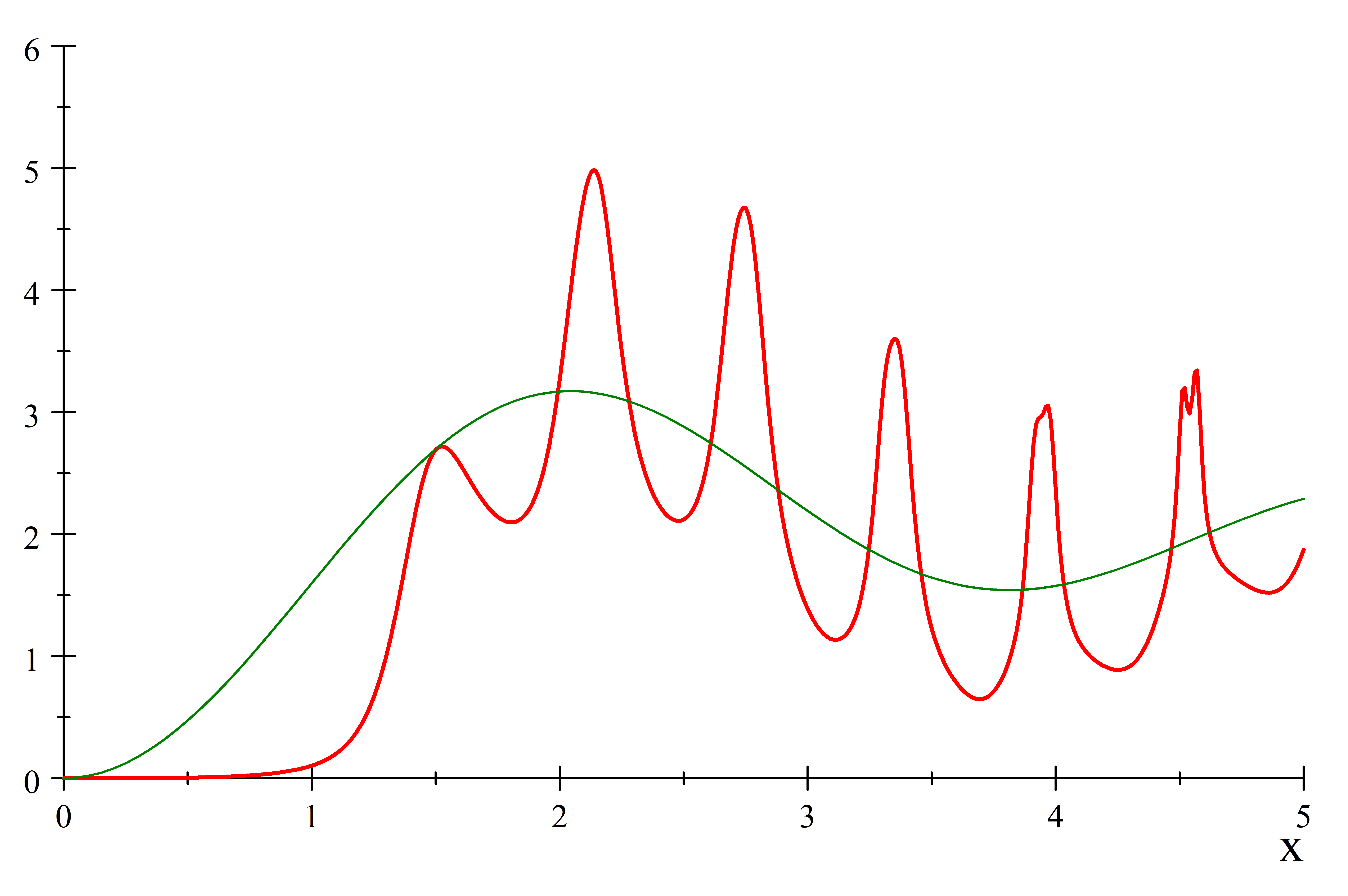}
	   \caption{$l=1$ to $6$ summed contributions to $\sigma_{TE}+\sigma_{TM}$ \\ with $\sigma_{P}$ in green.}
	\end{minipage}
	\vspace{5mm}
	\begin{minipage}[c]{0.75\textwidth}
		\centering
		\includegraphics[height=7cm]{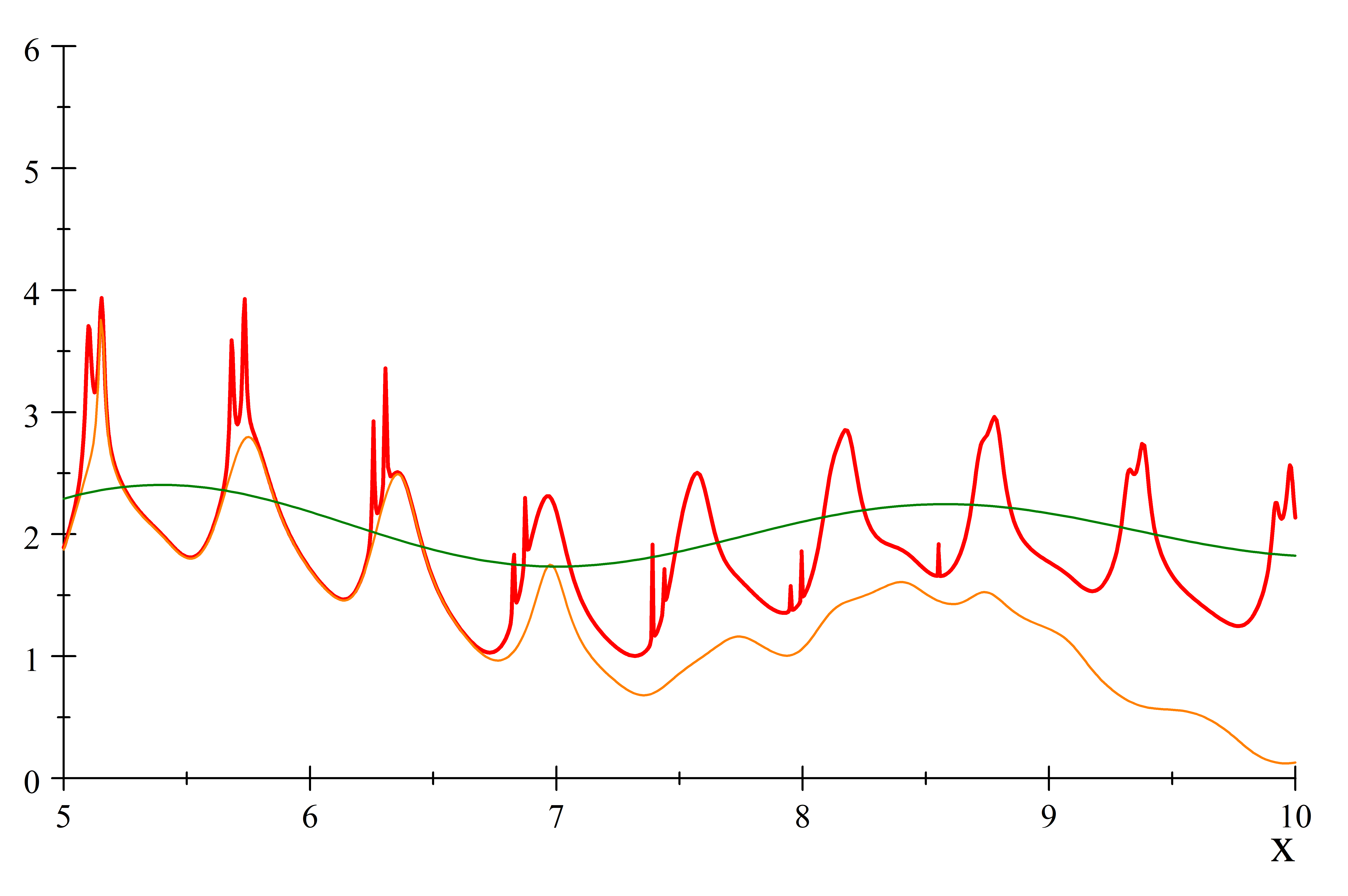}
	\caption{$l=1$ to $12$ summed contributions to $\sigma_{TE}+\sigma_{TM}$ in red and
		$\sigma_{P}$ in green ($l=1$ to $6$ summed contributions in
		orange).}
	\end{minipage}
\end{figure}

\end{document}